\documentclass[letterpaper]{article} % DO NOT CHANGE THIS
\usepackage{aaai2026}  % DO NOT CHANGE THIS
\usepackage{times}  % DO NOT CHANGE THIS
\usepackage{helvet}  % DO NOT CHANGE THIS
\usepackage{courier}  % DO NOT CHANGE THIS
\usepackage[hyphens]{url}  % DO NOT CHANGE THIS
\usepackage{graphicx} % DO NOT CHANGE THIS
\usepackage{natbib}  % DO NOT CHANGE THIS AND DO NOT ADD ANY OPTIONS TO IT
\usepackage{caption} % DO NOT CHANGE THIS AND DO NOT ADD ANY OPTIONS TO IT
\usepackage{algorithm, algpseudocode}

\usepackage{newfloat}
\usepackage{listings}
\DeclareCaptionStyle{ruled}{labelfont=normalfont,labelsep=colon,strut=off} % DO NOT CHANGE THIS
\floatstyle{ruled}
\newfloat{listing}{tb}{lst}{}
\floatname{listing}{Listing}
\usepackage[table, dvipsnames]{xcolor}
\usepackage{booktabs,multirow,array}
\usepackage[T1]{fontenc}
\usepackage{amsmath}
\usepackage{bm}
\usepackage{physics}
\usepackage{subcaption}
\usepackage{mathtools}
\usepackage{pifont}
\usepackage{etoolbox}
\usepackage{xspace}
\usepackage{amssymb}

\usepackage{silence}
\newcommand{\rv}[1]{{\boldsymbol{\mathbf{#1}}}}
\newcommand{\vaeenc}{q_{\rv{\psi}}(\rv{z} \vert \rv{x})}
\newcommand{\vaedec}{p_{\rv{\zeta}}(\rv{x} \vert \rv{z})}
\newcommand{\vaeprior}{p(\rv{z})}
\newcommand{\ours}{M\textsuperscript{3}ashy\xspace}

\definecolor{airforceblue}{rgb}{.3,.3,.8}

\AtEndPreamble{
    \usepackage[capitalize]{cleveref}
    \crefname{section}{Sec.}{Secs.}
    \Crefname{section}{Section}{Sections}
    \Crefname{table}{Table}{Tables}
    \crefname{table}{Tab.}{Tabs.}
    \Crefname{algorithm}{Algorithm}{Algorithms}
    \crefname{algorithm}{Alg.}{Algs.}
    \Crefname{figure}{Figure}{Figures}
    \crefname{figure}{Fig.}{Figs.}
}

\RequirePackage{enumitem}
\setlist[itemize]{noitemsep,leftmargin=*,topsep=0em}
\setlist[enumerate]{noitemsep,leftmargin=*,topsep=0em}

\makeatletter
\DeclareRobustCommand\onedot{\futurelet\@let@token\@onedot}
\def\@onedot{\ifx\@let@token.\else.\null\fi\xspace}

\def\eg{\emph{e.g}\onedot} 
\def\ie{\emph{i.e}\onedot} 
 
\def\etc{\emph{etc}\onedot}

\makeatother

\definecolor{iccvblue}{rgb}{0.21,0.49,0.74}
\newcommand{\iccvblue}[1]{{\color{iccvblue}#1}}
\definecolor{LightCyan}{rgb}{0.93,1,1}

\usepackage{algpseudocode}

\title{\ours{}: \iccvblue{M}ulti-\iccvblue{M}odal \iccvblue{Ma}terial \iccvblue{S}ynthesis via \iccvblue{Hy}perdiffusion}

\author {
    Chenliang Zhou, Zheyuan Hu, Alejandro Sztrajman, Yancheng Cai, Yaru Liu, Cengiz Oztireli
}
\affiliations{
    Department of Computer Science and Technology, University of Cambridge\\
    chenliang.zhou@cst.cam.ac.uk
}
\begin{document}

\twocolumn[{%
\renewcommand\twocolumn[1][]{#1}%
\maketitle
\vspace{-5em}
\centering
\includegraphics[width=\linewidth]{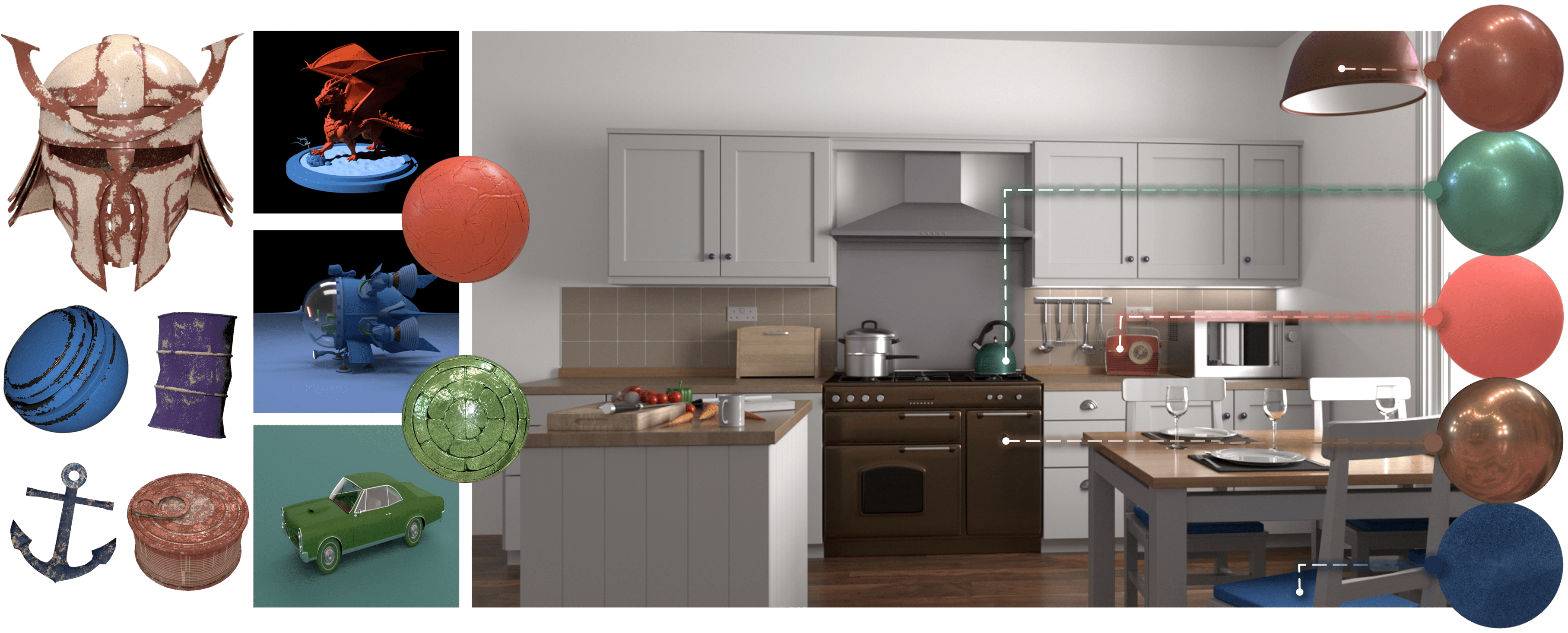}
% \vspace{-2em}
\captionof{figure}{3D models and scenes rendered with our synthesized neural materials demonstrate visually rich results.\vspace{1em}}
\label{fig:teaser}
}]

\begin{abstract}
High-quality material synthesis is essential for replicating complex surface properties to create realistic scenes. Despite advances in the generation of material appearance based on analytic models, the synthesis of real-world measured BRDFs remains largely unexplored. To address this challenge, we propose \ours, a novel \textbf{m}ulti-\textbf{m}odal \textbf{ma}terial \textbf{s}ynthesis framework based on \textbf{hy}perdiffusion. \ours{} enables high-quality reconstruction of complex real-world materials by leveraging neural fields as a compact continuous representation of BRDFs. Furthermore, our multi-modal conditional hyperdiffusion model allows for flexible material synthesis conditioned on material type, natural language descriptions, or reference images, providing greater user control over material generation. To support future research, we contribute two new material datasets and introduce two BRDF distributional metrics for more rigorous evaluation. We demonstrate the effectiveness of \ours through extensive experiments, including a novel statistics-based constrained synthesis, which enables the generation of materials of desired categories.
% Extensive experiments demonstrate the effectiveness of \ours, including a novel constrained synthesis method that enables the generation of materials with specific statistical properties. To support further research, we also contribute two new datasets: AugMERL, an augmented collection of measured BRDFs, and NeuMERL, a dataset of BRDFs represented through implicit neural representations. 
\end{abstract}

\section{Introduction}
Material synthesis plays a crucial role in visual computing, enabling the creation of realistic material appearances for applications in scene understanding~\cite{gupta2013perceptual}, material recognition~\cite{bell2015material}, intrinsic image decomposition~\cite{bousseau2009intrinsic}, generative image synthesis~\cite{karras2019style}, and physics-based vision for simulation~\cite{wu2017marrnet}. Material appearance is commonly modeled by the \emph{bidirectional reflectance distribution function~(BRDF)}. Although there has been significant progress~\cite[\eg,][]{gatys2015texture, zhou2018non} in generative modeling of analytic BRDFs~\cite{phong, ggx}, there is a lack of work focusing on that of measured ones. For the analytic BRDFs, the actual per-point reflectance model is typically relatively simple and low-dimensional~\cite{ngan2005experimental, guarnera2016}. In contrast, a measured BRDF is tabulated from real-world capture and can be substantially higher-dimensional~\cite[\eg,][]{Matusik2003datadriven}, with the ability to represent complex and irregular scattering behaviors that exceed the expressiveness of analytic models~\cite{ngan2005experimental}.
However, the high-dimensionality often hinders the performance of learning-based methods.
To address this gap, we propose \ours, a framework for realistic material generation that leverages neural fields~\cite{sztrajman2021nbrdf} as an alternative low-dimensional continuous representation for material appearance that combines high-quality reconstruction with memory efficiency. This simplifies the learning process, enabling the model to capture the underlying material distribution more efficiently.

An additional challenge in material synthesis is the lack of robust quantitative metrics for evaluating synthesis quality, making it difficult to assess and compare different approaches, unlike generative models in other fields~\cite{theis2016evalgen, betzalel2022evalgen}. A final limitation is the absence of multi-modal conditioning, which would enable users to guide the synthesis process using diverse inputs, such as material type, text descriptions or reference images. This limitation reduces the flexibility and control available to artists and designers. To address these limitations, we propose a set of novel BRDF distributional metrics and leverage a multi-modal conditional hyperdiffusion model to support flexible user input.

The main contributions of this work are as follows:
\begin{itemize}
      \item A novel material synthesis pipeline using a multi-modal conditional hyperdiffusion model that supports user-specified material generation via material type, natural language descriptions, or image references.
      \item A thorough evaluation of \ours{}’s effectiveness, including a set of novel BRDF distributional metrics and a novel constrained synthesis experiment to synthesize materials of desired categories.
	  \item Two new datasets: AugMERL, an enhanced collection of tabulated BRDF values, and NeuMERL, a dataset of materials represented through INRs.
\end{itemize}

%--------------------------------------------------------------------------
\section{Related Work}
\paragraph{Material modeling}
Material appearance has been widely modeled by the \emph{bidirectional reflectance distribution function~(BRDF)}~\cite{Nicodemus1977GeometricalCA, guarnera2016, montes2012overview, ngan2005experimental, westin2004comparison}.
% , which encodes the complex light-surface interaction~\cite{Guarnera2016BRDFsurvey, montes2012overview, ngan2005experimental, westin2004comparison}.
While analytic BRDF models~\cite{phong, cook-torrance, ggx, disney} offer efficient reconstruction and editing, their simplified assumptions limit the representation of complex real-world materials~\cite{ngan2005experimental, guarnera2016,remapping2017,remapping2019}. Data-driven approaches offer higher realism~\cite{Matusik2003datadriven}, although they are often hard to manipulate and require large storage. Dimensionality reduction techniques can alleviate this issue, but at the expense of compromising material quality~\cite{lawrence2004efficient, NielsenPCA2015}. Recently, deep learning methods provide efficient, low-dimensional representations~\cite{deepbrdf, zheng, sztrajman2021nbrdf, gokbudak2023hypernetworks, Fanlayered, metalayered}. Our work leverages a neural field architecture~\cite{sztrajman2021nbrdf} for efficient, realistic material modeling. Please see \cref{sec:Related Work on Material Acquisition and Databases} for additional related work on material acquisition and databases.

\paragraph{Material synthesis}
Material synthesis based on analytic models has been widely explored~\cite{zhang2024textmat, memery2023materialNLP, Chen_2023_ICCV, Chen2023Text2TexTT, hu2023generating, tchapmi2022generating, xu23MATLABER, Henzler2021GenBRDFimg}. For data-driven representations, previous works have resourced to various strategies, including dimensionality reduction~\cite{Matusik2003datadriven, abdi2010principal}, perceptual mappings~\cite{NielsenPCA2015, serrano2016, sun2018connecting} and deep learning~\cite{deepbrdf, gokbudak2023hypernetworks}, but these approaches do not offer a generative modeling of materials.
% Matusik~\etal~\cite{Matusik2003datadriven} pioneered data-driven BRDF generation with principal component analysis (PCA)~\cite{abdi2010principal}, enabling BRDF creation through interpolation in a reduced space. Nielsen~\etal~\cite{nielsen2015} improved PCA-based generation by introducing a log-relative mapping of BRDF data, and applied it to the problem of optimal sparse reconstruction of BRDFs. Other works have used PCA-based workflows with log-relative mapping for different tasks, such as BRDF data augmentation~\cite{serrano2016} and diffuse/specular term separation~\cite{sun2018connecting}.  
\citet{deepbrdf} use a convolutional autoencoder to learn a low-dimensional manifold from a measured BRDF database, enabling material editing. However, their BRDF representation is constrained to a fixed resolution with high storage requirements. \citet{gokbudak2023hypernetworks} leverage a hypernetwork architecture to predict the weights of a neural fields representation of material appearance. Nevertheless, their approach requires sample measurements of BRDF data as input and is thus limited to sparse reconstruction. There also exist methods for text-~\cite{xu23MATLABER, memery2023materialNLP} and image-conditioned synthesis~\cite{deepbrdf, Henzler2021GenBRDFimg}, but these either focus on analytic materials or do not provide a generative modeling approach.

In this work, we introduce a generative approach for measured real-world material based on a multi-modal hyperdiffusion architecture. Our method generates continuous, low-dimensional representations of materials, and can be conditioned on material type, natural language, and images. In \cref{tbl:cf_prior_art}, we compare various state-of-the-art material modeling methods: Our approach is the first generative pipeline to support unconditional, multi-modal, and constrained synthesis of measured real-world materials while also contributing novel datasets and introducing quantitative metrics for material synthesis evaluation.
%\footnote{DeepBRDF \cite{deepbrdf}, \cite{memery2023materialNLP} via the NVIDIA Omniverse parametric material, \cite{Henzler2021GenBRDFimg} via a parametric model from flash images, \cite{xu23MATLABER} via the Cook-Tolerance model.}

\begin{table*}
    \centering
    % \setlength{\tabcolsep}{1mm}
    % \resizebox{\linewidth}{!}{
    \begin{tabular}{lcccccccc}
        \toprule
        BRDF modeling method         & Measured   & Generative & Type & Text & Image & CS & Datasets & Metrics \\ 
        \midrule
        DeepBRDF~\cite{deepbrdf} & \ding{51}           & \ding{55}  & \ding{55}     & \ding{55}     & \ding{51} & \ding{55}  & \ding{55} & \ding{55}             \\  
        \citet{Henzler2021GenBRDFimg} & \ding{55}          & \ding{51}  & \ding{55}     & \ding{55}     & \ding{51}  & \ding{55}  & \ding{55} & \ding{51}         \\  
        MATLABER~\cite{xu23MATLABER} & \ding{55}              & \ding{51} & \ding{55}      & \ding{51} & \ding{55}    & \ding{55}  & \ding{55}  & \ding{55}             \\ 
        \citet{memery2023materialNLP} & \ding{55}          & \ding{51}    & \ding{55}   & \ding{51} & \ding{55}   & \ding{55}  & \ding{55}   & \ding{55}             \\  
        \citet{gokbudak2023hypernetworks} & \ding{51} & \ding{55}  & \ding{55}     & \ding{55}     & \ding{51}  & \ding{55}  & \ding{55} & \ding{55}   \\ 
        \ours (ours)              & \ding{51} & \ding{51}   & \ding{51} & \ding{51} & \ding{51} & \ding{51}  & \ding{51} & \ding{51}         \\
        \bottomrule
    \end{tabular}
    % }
    \caption{Comparison of material modeling methods. Our \ours is the first generative pipeline for measured real-world materials that supports both unconditional and multi-modal conditional synthesis guided by type, text, or image. It also enables a statistics-based constrained synthesis (CS) and introduces novel datasets and material distributional metrics.}
    \label{tbl:cf_prior_art}
\end{table*}

\section{Methods}
\begin{figure}
      \centering
       \includegraphics[width=\linewidth]{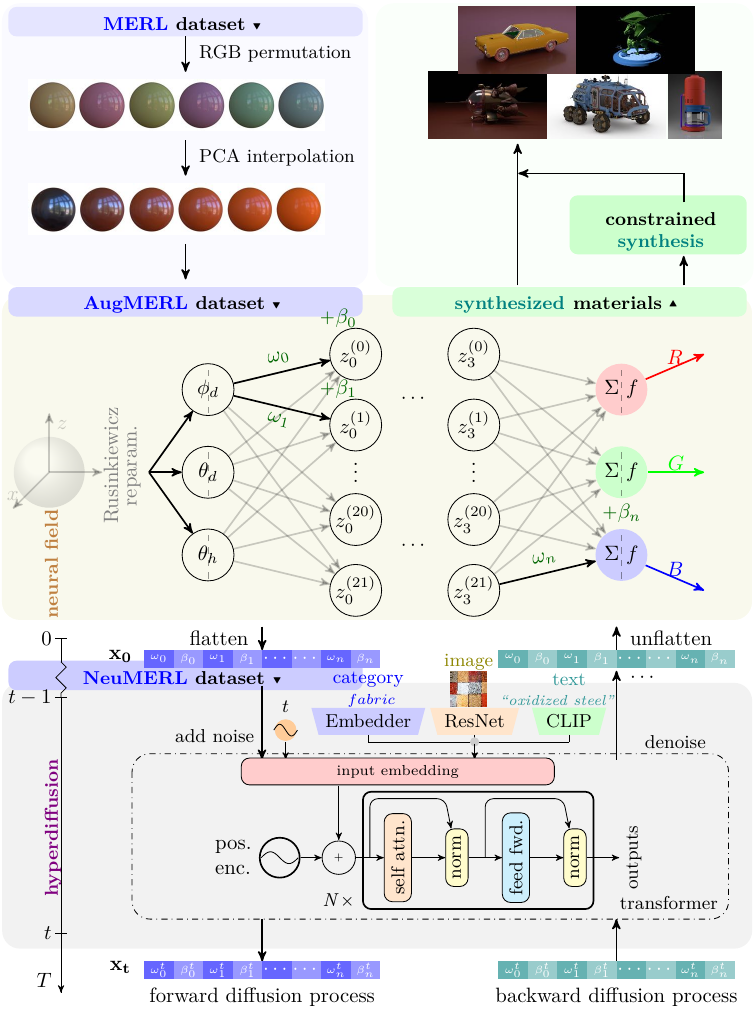}
      \caption{An overview of \ours, our novel neural material synthesis framework, consisting of three main stages. 1 (top left): Data augmentation using RGB permutation and PCA interpolation to create an expanded dataset, \emph{AugMERL}; 2 (middle): Neural field fitted to individual materials, resulting in \emph{NeuMERL}, a dataset of neural material representations; and 3 (bottom): Training a multi-modal conditional hyperdiffusion on NeuMERL to enable conditional synthesis of high-quality, diverse materials guided by inputs such as material type, text descriptions, or reference images. We further propose a novel statistics-based constrained synthesis method to generate materials of a specified type (top right).}
      \label{fig:pipeline_overview}
\end{figure}

An overview of our material synthesis pipeline, \ours, is shown in \cref{fig:pipeline_overview}, consisting of three main stages. First, we augment the MERL dataset through RGB permutation and PCA interpolation, generating the \emph{Augmented MERL (AugMERL)} dataset (\cref{sec:Data Augmentation}). Next, we adopt neural fields as a low-dimensional, continuous representation for materials, fitting them to individual materials in AugMERL to create a new dataset of neural material representations, \emph{Neural MERL (NeuMERL)}. Finally, we train a transformer-based, multi-modal hyperdiffusion model on NeuMERL to capture the complex distribution of neural materials, enabling high-fidelity and diverse synthesis through unconditional, multi-modal conditional, and constrained generation (\cref{sec:Experiments}).

\subsection{Data Augmentation}
\label{sec:Data Augmentation}
We utilize the MERL dataset~\cite{Matusik2003datadriven}, which includes 100 materials, each represented by $D_\text{MERL}=90 \times 90 \times 180$ densely sampled BRDF values.

\begin{figure}[tbh]
  \centering
  \begin{subfigure}{0.075\textwidth}
      \includegraphics[width=\textwidth]{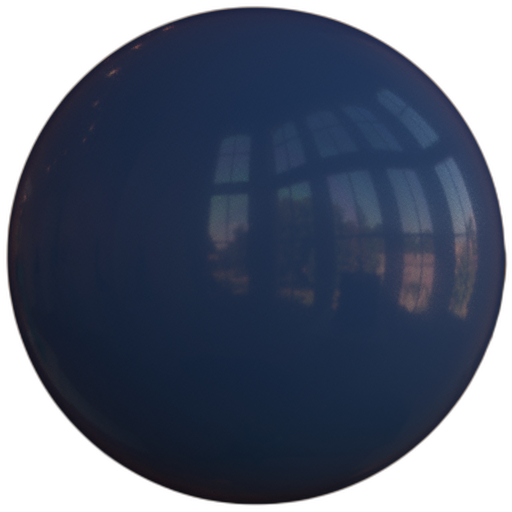}
      \caption*{RGB}
  \end{subfigure}
  \begin{subfigure}{0.075\textwidth}
      \includegraphics[width=\textwidth]{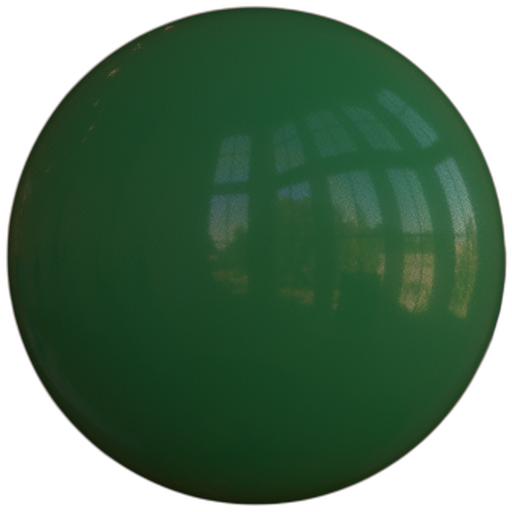}
      \caption*{RBG}
  \end{subfigure}
    \begin{subfigure}{0.075\textwidth}
      \includegraphics[width=\textwidth]{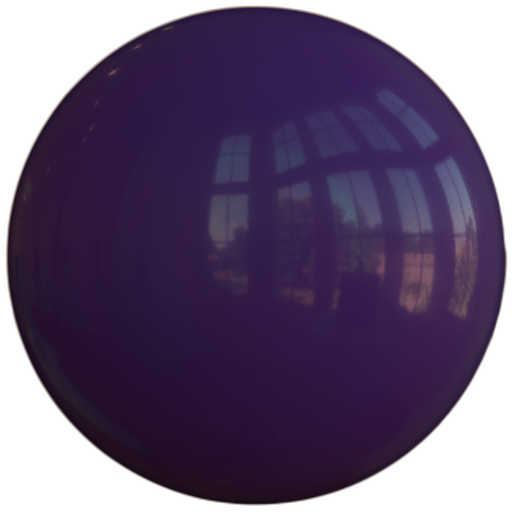}
      \caption*{GRB}
  \end{subfigure}
    \begin{subfigure}{0.075\textwidth}
      \includegraphics[width=\textwidth]{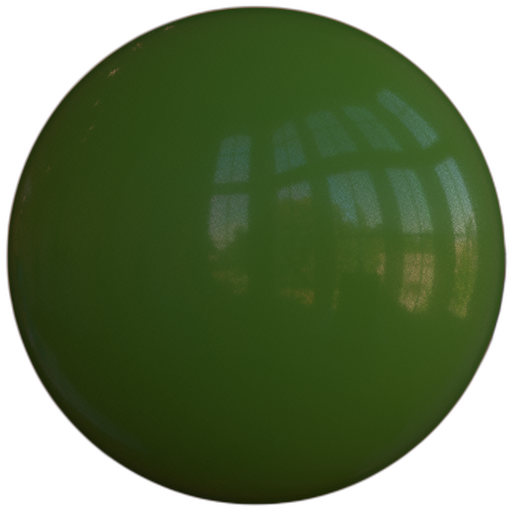}
      \caption*{GBR}
  \end{subfigure}
    \begin{subfigure}{0.075\textwidth}
      \includegraphics[width=\textwidth]{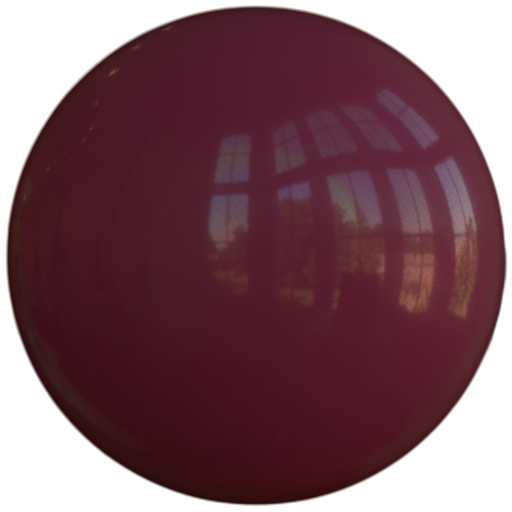}
      \caption*{BRG}
  \end{subfigure}
    \begin{subfigure}{0.075\textwidth}
      \includegraphics[width=\textwidth]{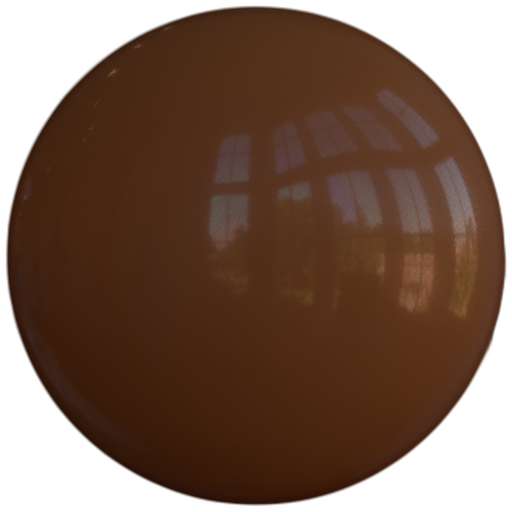}
      \caption*{BGR}
  \end{subfigure}
  \caption{Six RGB permutations of the MERL material \emph{blue acrylic}. (a) represents the original material. This permutation strategy expands the dataset by a factor of 6.}
  \label{fig:rgb-permutation}
\end{figure}

Through experimentation, we determined that 100 samples are insufficient for effective hyperdiffusion training. To address this, we augment the MERL dataset using RGB permutation and PCA interpolation. First, we permute the three color channels (RGB) of each MERL sample, yielding an expanded dataset of $100 \times 3! = 600$ samples. An example of RGB permutation is illustrated in \cref{fig:rgb-permutation}.

\begin{figure}[tbh]
  \centering
  \begin{subfigure}{0.075\textwidth}
      \includegraphics[width=\textwidth]{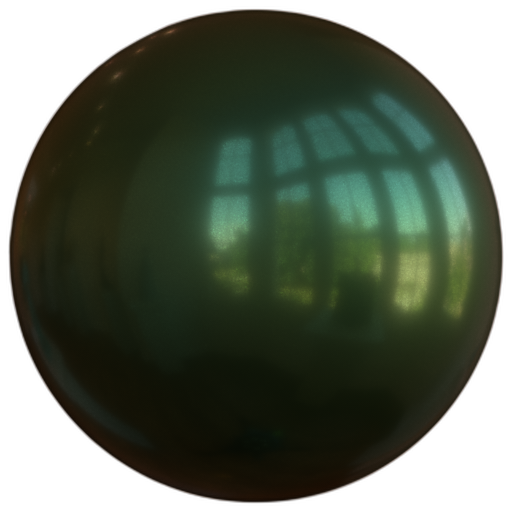}
      \caption*{0\%}
  \end{subfigure}
  \begin{subfigure}{0.075\textwidth}
      \includegraphics[width=\textwidth]{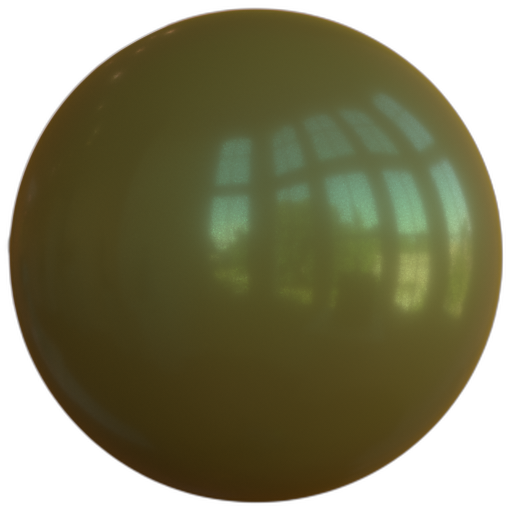}
      \caption*{20\%}
  \end{subfigure}
    \begin{subfigure}{0.075\textwidth}
      \includegraphics[width=\textwidth]{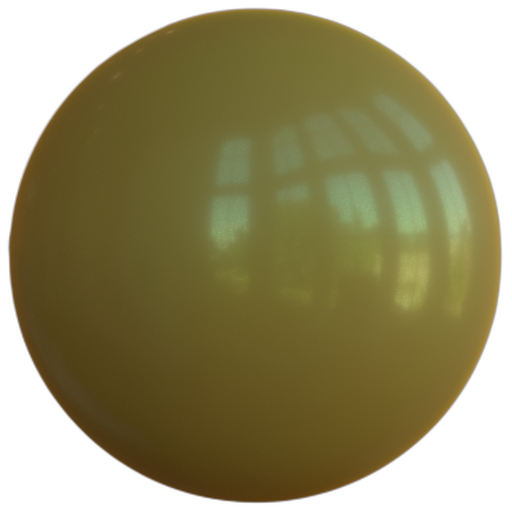}
      \caption*{40\%}
  \end{subfigure}
    \begin{subfigure}{0.075\textwidth}
      \includegraphics[width=\textwidth]{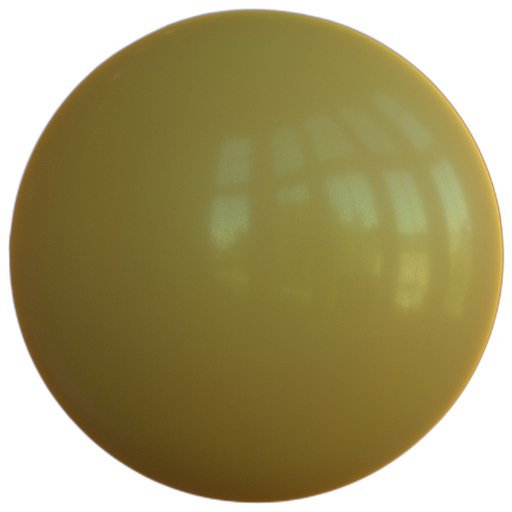}
      \caption*{60\%}
  \end{subfigure}
    \begin{subfigure}{0.075\textwidth}
      \includegraphics[width=\textwidth]{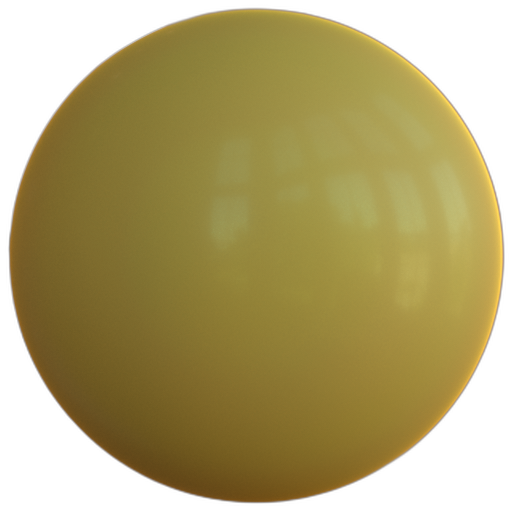}
      \caption*{80\%}
  \end{subfigure}
    \begin{subfigure}{0.075\textwidth}
      \includegraphics[width=\textwidth]{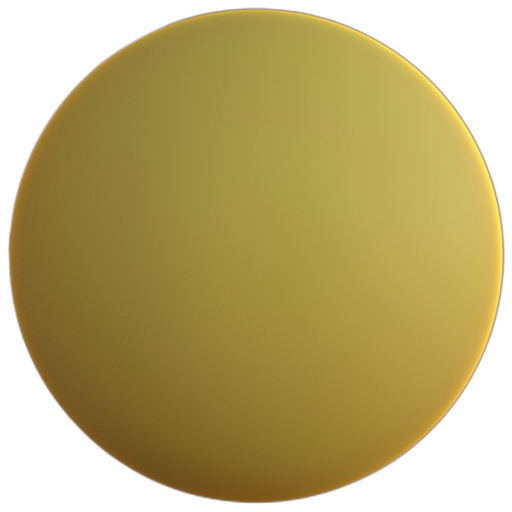}
      \caption*{100\%}
  \end{subfigure}
  \caption{Linear interpolation of two MERL materials, (a) \emph{green metallic paint} and (f) \emph{yellow plastic}, in the PCA space.}
  \label{fig:pca-interpolation}
\end{figure}

After applying RGB permutation, we perform principal component analysis (PCA)~\cite{abdi2010principal} to reduce the dimensionality of the BRDF data from \( D_\text{MERL} \) to 300. In this lower-dimensional space, we perform linear interpolation to further augment the dataset, expanding it to \( 2400 \) materials. Compared to direct linear interpolation in the high-dimensional BRDF space, interpolation in PCA space is more effective in capturing the underlying structure of the BRDF data and yields perceptually accurate results~\cite{Matusik2003datadriven, lawrence2004efficient, romeiro2010blind}. An example of materials generated through PCA interpolation is shown in \cref{fig:pca-interpolation}.
We refer to this augmented dataset as \emph{Augmented MERL (AugMERL)}.
% denoted $\mathcal{D}_\text{AugMERL} \in \mathbb{R}^{N \times D_\text{MERL}}$.
For additional details on PCA, please see \cref{sec:principal-component-analysis} in the supplementary.

\subsection{Neural Field Fitting}
\label{sec:NBRDF Fitting}
Neural fields provide a low-dimensional, continuous representation for material data. Following prior work~\cite{sztrajman2021nbrdf}, we overfit a compact neural field $f_r^{\rv{\xi}}$ parameterized by $\rv{\xi}$, to each material in AugMERL. Once fitted, we treat the flattened weights of $f_r^{\rv{\xi}}$ as the material's neural representation. With 2400 materials in AugMERL, this process yields a dataset of 2400 neural material representations, which we refer to as \emph{Neural MERL (NeuMERL)}.
% , denoted $\mathcal{D}_\text{NeuMERL} \in \mathbb{R}^{N \times D_\text{NF}}$, where $D_\text{NF}$ is the sum of dimensionalities of the weights in our neural field $f_r^\rv{\xi}$.

Representing materials as flattened 1D vectors enables a flexible framework for modeling complex distributions, abstracting away the underlying data’s dimensionality. This approach makes our pipeline adaptable to diverse data formats. In the following section, we detail three key techniques employed in fitting the neural fields.

\paragraph{Rusinkiewicz reparametrization} In our preliminary experiments and other studies~\cite[\eg,][]{sztrajman2021nbrdf, zhou2024physically}, it is observed that directly using the conventional BRDF input format -- namely the incident and outgoing directions $\rv{\omega}_i, \rv{\omega}_o \in \mathbb{R}^3$ -- can complicate the fitting process for certain materials and occasionally introduce undesirable artifacts possibly due to the high dimensionality of the input.
% , such as tangential discontinuities. These artifacts may stem from violations of Helmholtz reciprocity~\cite{stokes1849perfect} when neural fields are used to model BRDFs.
To address this, we employ the Rusinkiewicz reparametrization~\cite{rusinkiewicz1998}, which defines the half and difference vectors $\rv{h}$ and $\rv{d}$ as follows:
\begin{equation}
    \rv{h} \coloneqq \frac{\rv{\omega}_i + \rv{\omega}_o}{\lVert \rv{\omega}_i + \rv{\omega}_o \rVert};\,\,
    \rv{d} \coloneqq R_{\hat{\rv{b}},-\theta_\rv{h}} R_{\hat{\rv{n}},-\varphi_\rv{h}} \rv{\omega}_i,
\end{equation}
where $R_{\rv{v}, \alpha}$ denotes a rotation around the vector $\rv{v}$ by the angle $\alpha$, $\hat{\rv{n}}$ is the surface normal, and $\hat{\rv{b}}$ is the surface binormal. This reparametrization helps improve the robustness of the neural field fitting process by addressing reciprocity constraints more directly~\cite{sztrajman2021nbrdf, zhou2024physically}.

We then proceed by adopting the spherical coordinates of the half and difference vectors $\rv{h}$ and $\rv{d}$, specifically $\theta_{\rv{h}}, \varphi_{\rv{h}}, \theta_{\rv{d}}, \varphi_{\rv{d}}$, as inputs to our neural fields. A further advantage of using the Rusinkiewicz reparametrization is that, since our materials are isotropic, the BRDF remains invariant with respect to $\varphi_{\rv{h}}$. Consequently, we can omit this parameter, reducing the input complexity from $(\rv{\omega}_i, \rv{\omega}_o) \in \mathbb{R}^6$ to $(\theta_{\rv{h}}, \theta_{\rv{d}}, \varphi_{\rv{d}}) \in [0, \frac{\pi}{2}]^2 \times [0, \pi)$. This reparametrization enhances the efficiency of our neural field representation without sacrificing accuracy.

\paragraph{Mean absolute logarithmic loss} The high dynamic range of BRDF values makes fitting reflectance data particularly sensitive to error distribution. For low reflectance values, even minor fitting errors can have a large impact on the loss, causing shifts in perceived ``hue'' in rendered images, which leads to unrealistic colors and reduced visual fidelity. To address these issues, we employ a mean absolute logarithmic loss for BRDF values~\cite{sztrajman2021nbrdf}:
% \begin{equation}
% % \small
%     \mathcal{L}_\text{NF}(\rv{\xi}) \coloneqq \underset{\theta_{\rv{h}}, \theta_{\rv{d}}, \varphi_{\rv{d}}}{\mathbb{E}} \Big[ \left| \log \left(1 + f_r \cos\theta_i \right) - \log \left(1 + f_r^{\rv{\xi}} \cos\theta_i \right) \right| \Big],
%     % \label{eq:nbrdf-loss}
% \end{equation}
\begin{equation}
    \mathcal{L}_\text{NF}(\rv{\xi}) \coloneqq \mathbb{E}_{
        \raisebox{-4mm}{
            $\mathclap{
                \substack{
                    \theta_{\rv{h}}, \theta_{\rv{d}}, \varphi_{\rv{d}}
                }
            }$
        }
    }\mkern-5mu \Big[ \left| \log \left(1 + f_r \cos\theta_i \right) - \log \left(1 + f_r^{\rv{\xi}} \cos\theta_i \right) \right| \Big]
    % \label{eq:nbrdf-loss}
\end{equation}
where \( f_r \) denotes the ground-truth BRDF, and \( \theta_i \) is the polar angle of the incident direction. This loss is computed per color channel, offering a balanced approach that stabilizes training across samples with both low and high values. Consequently, it enhances the model’s capability to manage dynamic reflectance variations, leading to more realistic color reproduction and improved visual fidelity.

\paragraph{Weight initialization} Ideally, the neural materials in NeuMERL should originate from a consistent distribution. However, due to a phenomenon known as \emph{weight symmetry}~\cite{liao2016important}, we observe that different weights can yield the same neural field. For instance, swapping weights between two neurons in a hidden layer, or flipping the signs of both input and output weights for a neuron before an odd, linear, or piecewise-linear activation function like ReLU~\cite{nair2010rectified}, results in an identical neural field. To address this, we propose using the optimized weights from the first fitted neural field as the initialization for subsequent neural field fittings. This approach helps align the weights across all fitted fields, promoting consistency within NeuMERL and facilitating smoother training of the hyperdiffusion model in the next stage.

\subsection{Multi-Modal Conditional Hyperdiffusion}
\label{sec:Hyperdiffusion on NBRDF Weights}
To model the complex distribution within the NeuMERL dataset, we utilize a diffusion process~\cite{ho2020denoisingdiffusion, peebles2022learning, erkoc2023hyperdiffusion}. Specifically, we employ a transformer-based denoising network~\cite{vaswani2017attention}, leveraging its demonstrated efficacy~\cite{peebles2022learning} and its attention mechanisms allowing for an effective focus on relevant information, enhancing the network’s ability to capture intricate dependencies in the data.

Our hyperdiffusion supports conditioning across three modalities: material type (represented as integers), text description, and reference images. We utilize a categorical encoding for material type, an augmented CLIP text embedding~\cite{radford2021learning, zhou2023clip} for text, and a ResNet~\cite{he2016deep} for images. This multi-modal conditioning approach enables material synthesis to be guided by different user inputs, enhancing workflow intuitiveness and accessibility and allowing for a more smooth and accurate translation of creative vision into generated materials. For conditional sampling, we employ classifier-free guidance (CFG)~\cite{ho2022cfg}.

% The forward and backward processes in hyperdiffusion are modeled as Markov chains with a total timestep $T$. To train a learnable model \(\rv{\epsilon}_\rv{\eta}(\rv{x}_t, t, \rv{y})\) parameterized by \(\rv{\eta}\), we minimize the score matching objective:
% \begin{equation}
%     \mathcal{L}_\text{HD}(\rv{\eta}) \coloneqq \mathbb{E}_{\rv{x}_0,t \sim \mathcal{U}(1,T), \rv{\epsilon} \sim \mathcal{N}(0, I)} \left[ \| \rv{\epsilon} - \rv{\epsilon}_\rv{\eta}(\rv{x}_t, t, \rv{y}) \|_2^2 \right],
% \end{equation}
% where \(\mathcal{U}(1,T)\) represents the uniform distribution over \(\{1,2,\ldots,T\}\) and $\rv{y}$ is the conditional input embedding. This objective encourages the model to accurately predict the noise \(\rv{\epsilon}\), effectively guiding the denoising process.

% During inference, the network enables sampling via an iterative process~\cite{song2022ddim}, leveraging the factorization of the learned distribution as
% \begin{equation}
% \label{eq:ddpm-gen}
%     p_\rv{\eta}(\rv{x}) = p(\rv{x}_T) p_\rv{\eta}(\rv{x}_0 \vert \rv{x}_T) = p(\rv{x}_T) \prod_{t=1}^T p_\rv{\eta} (\rv{x}_{t-1} \vert \rv{x}_t)
% \end{equation}
% for $p(\rv{x}_T) \coloneqq \mathcal{N}(0, I)$. For conditional sampling, we employ classifier-free guidance (CFG)~\cite{ho2022cfg}.

Please refer to \cref{sec:more-diffusion-model,sec:Transformer Backbone in Hyperdiffusion,sec:attention-module,sec:Conditional Sampling with Classifier-Free Guidance} in the supplementary for further details on the diffusion model, transformer, attention mechanism, and CFG, respectively.
\section{Experiments}
\label{sec:Experiments}
\begin{figure*}[tbh]
\centering
\includegraphics[width=0.95\linewidth]{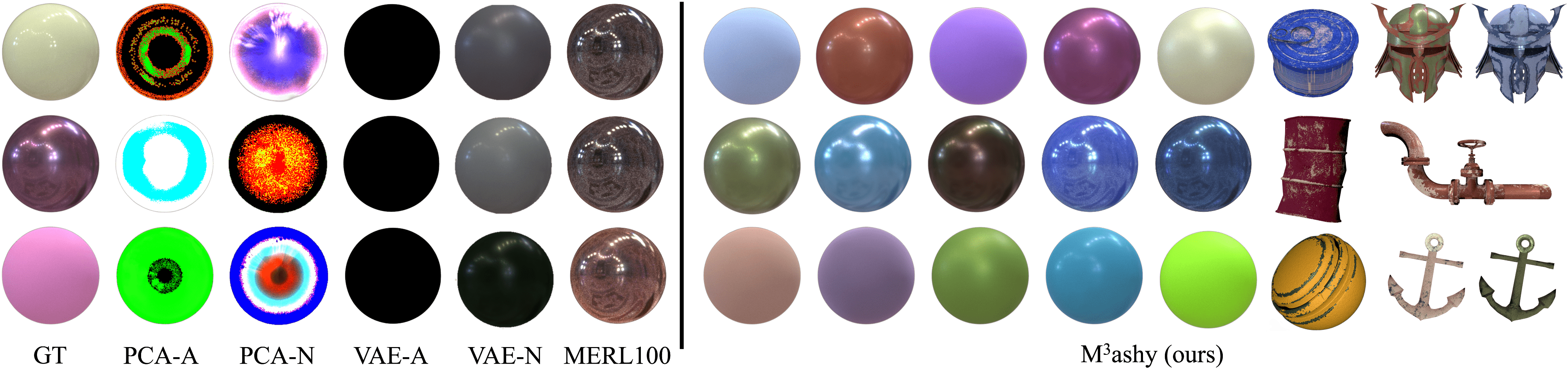}
\caption{Material synthesis. Baseline models fail to capture the underlying distribution effectively, resulting in homogeneous outputs or severe artifacts. In contrast, \ours successfully captures the complex neural material distribution, achieving significantly better fidelity and diversity. Our materials also support spatially varying rendering configurations (last three columns).}
\label{fig:uncond-spheres}
\end{figure*}
In this section, we present extensive experiments on \ours.
% , beginning with an overview of our datasets~(\cref{sec:exp-datasets}) and introducing novel material distributional metrics~(\cref{sec:performance-metrics}). We then demonstrate the effectiveness of our pipeline through experiments on unconditional synthesis~(\cref{sec:unconditional-synthesis}), multi-modal conditional synthesis~(\cref{sec:conditional-synthesis}), and constrained synthesis~(\cref{sec:Constrained Synthesis}).
For additional details, please refer to the appendix: model and experiment specifics in \cref{sec:Model Details,sec:Experiment Details}, further results in \cref{sec:Further Results}, and more experiments in \cref{sec:Additional Experiments}.

\subsection{Dataset}
\label{sec:exp-datasets}
We fit neural fields to individual materials in the AugMERL dataset (\cref{sec:Data Augmentation}), which is derived from the MERL BRDF dataset~\cite{Matusik2003datadriven}. Our hyperdiffusion model is trained on the NeuMERL dataset (\cref{sec:NBRDF Fitting}), which consists of neural material representations. The training-validation split is 80\%-20\% for each individual fitting on AugMERL and 95\%-5\% for NeuMERL. All samples derived from a given material remain within one split. In the constrained synthesis experiments in \cref{sec:Constrained Synthesis}, statistical information is gathered from the MERL dataset~\cite{Matusik2003datadriven}.

\subsection{Material Distributional Metrics}
\label{sec:performance-metrics}
We use Fréchet Inception Distance (FID)~\citep{heusel2017gans} as an image-based metric to assess the quality of rendered single-view images. % by comparing them to a reference set. 

% While image-based metrics can provide insight into synthesized materials by evaluating renderings from specific viewpoints, they have limitations. These metrics are heavily influenced by lighting and viewing conditions, which may obscure key differences in BRDF behavior and overlook important perceptual and physical characteristics~\cite{ngan2005experimental, Matusik2003datadriven, kim2011perception, lafortune1997non, kingma2013auto}.
To the best of our knowledge, effective metrics directly comparing material distributions are still lacking. Drawing inspiration from the metrics for point clouds~\cite{yang2019pointflow}, we introduce three novel material distributional metrics -- minimum matching distance (MMD), coverage (COV), and 1-nearest neighbor (1-NNA) -- to evaluate the fidelity and diversity of synthesized BRDF sets $\mathcal{S}$ relative to a reference set $\mathcal{R}$. Each metric is based on an underlying distance measure $d(f_r, f_r')$ between two BRDFs.

\paragraph{Minimum matching distance (MMD)} MMD measures the average distance from each reference BRDF to its nearest synthesized counterpart:
\begin{equation}
    \mathcal{L}_\text{MMD}^d(\mathcal{R}, \mathcal{S}) \coloneqq \frac{1}{\left|\mathcal{R} \right|} \sum_{f_r \in \mathcal{R}} \underset{f_r' \in \mathcal{S}}{\min} d(f_r, f_r')
\end{equation}
MMD evaluates the fidelity of the synthesized set relative to the reference, with a lower score indicating higher fidelity.

\paragraph{Coverage (COV)} COV calculates the proportion of reference BRDFs that are ``covered'' by the synthesized set. A reference BRDF is considered covered if it is the closest neighbor to at least one synthesized BRDF:
\begin{equation}
    \mathcal{L}_\text{COV}^d(\mathcal{R}, \mathcal{S}) = \frac{\left| \left\{ \underset{f_r \in \mathcal{R}}{\text{argmin }} d(f_r, f_r') \mid f_r' \in \mathcal{S} \right\} \right|}{\left|\mathcal{R} \right|}
\end{equation}
COV assesses the diversity of the synthesized set, with a higher score reflecting better coverage.

\paragraph{1-nearest neighbor (1-NNA)} 1-NNA is a leave-one-out metric that measures the similarity between the reference and synthesized BRDF distributions, capturing both diversity and fidelity:
{\small
\begin{equation}
    \mathcal{L}_\text{1-NNA}^d(\mathcal{R}, \mathcal{S}) \coloneqq \frac{\sum\limits_{f_r \in \mathcal{R}} \mathbb{I}\left[N_{f_r} \in \mathcal{R} \right] + \sum\limits_{f_r' \in \mathcal{S}} \mathbb{I}[N_{f_r'} \in \mathcal{S}]}{\left| \mathcal{S} \right| + \left| \mathcal{R} \right|},
\end{equation}}
where \(\mathbb{I}[\cdot]\) is the indicator function and \(N_{f_r}\) denotes the nearest neighbor of \(f_r\) in \((\mathcal{R} \cup \mathcal{S}) - \{f_r\}\). In this metric, each sample is classified as belonging to either the reference set \(\mathcal{R}\) or the synthesized set \(\mathcal{S}\) based on the membership of its nearest neighbor. If \(\mathcal{R}\) and \(\mathcal{S}\) are drawn from the same underlying distribution, the classifier's accuracy will approach 50\% with a large sample size.
% An accuracy closer to 50\% indicates greater similarity between \(\mathcal{R}\) and \(\mathcal{S}\), suggesting that the model has effectively learned the target distribution.

Each metric can be computed using an underlying distance \(d\). Potential options include rendering-based metrics such as root mean squared error (RMSE), peak signal-to-noise ratio (PSNR), and structural similarity index measure (SSIM)~\cite{SSIM04}.
% PSNR measures the reconstruction quality of images, while SSIM predicts perceived similarity.
Since higher values in PSNR and SSIM correspond to greater similarity, we negate these values -- resulting in NegPSNR and NegSSIM, respectively -- to make them plausible distance functions
% \footnote{Although NegPSNR and NegSSIM may not strictly meet the criteria for distance metrics (\eg, violating positivity), they suffice to provide useful relative distance measures for material distributional metrics.}.
To directly assess the distance between two BRDFs without relying on renderings, we also introduce the following BRDF L1 distance:
%, logarithmic (BRDF-L1-Log), and the neural field (BRDF-NF) distances (similar to \cref{eq:nbrdf-loss})
\begin{align}
    d_\text{BRDF-L1} \coloneqq \underset{\theta_{\rv{h}}, \theta_{\rv{d}}, \varphi_{\rv{d}}}{\mathbb{E}} \Big[ \left| f_r - f_r' \right| \Big];
%     &d_\text{BRDF-L1-Log} \coloneqq \underset{\theta_{\rv{h}}, \theta_{\rv{d}}, \varphi_{\rv{d}}}{\mathbb{E}} \Big[ \left| \log \left(1 + f_r \right) - \log \left(1 + f_r' \right) \right| \Big]; \label{eq:brdf-distance-l1-log} \\
%     &d_\text{BRDF-NF} \coloneqq \underset{\theta_{\rv{h}}, \theta_{\rv{d}}, \varphi_{\rv{d}}}{\mathbb{E}} \Big[ \left| \log \left(1 + f_r \cos\theta_i \right) - \log \left(1 + f_r' \cos\theta_i \right) \right| \Big].
\end{align}
% We append the name of the distance to the distributional metric to indicate the concrete combination, such as MMD-RMSE, COV-L1, 1-NNA-NF, \etc. 
For further background on the image-based metrics and validation of the proposed material distributional metrics, please refer to \cref{sec:Metric Details} in the supplementary.

\subsection{Unconditional Synthesis}
\label{sec:unconditional-synthesis}
We begin by presenting the results of unconditional synthesis of neural materials using \ours. 
% Given the novelty of our pipeline, we were unable to find suitable existing baselines for comparison. 
We compare our approach with the PCA-based method of \citet{NielsenPCA2015} and with the sparse reconstruction model of \citet{gokbudak2023hypernetworks}, which uses a hypernetwork to model measured tabular BRDF data. As \citeauthor{gokbudak2023hypernetworks}'s method is not generative, we extend it with a variational autoencoder (VAE) to enable comparison. Both methods are applied to the AugMERL (-A) and NeuMERL (-N) datasets, resulting in four baselines: VAE-A, VAE-N, PCA-A, and PCA-N.
% For synthesis, PCA-based baselines perform random linear interpolation in the principal space, while VAE-based baselines sample from a standard Gaussian prior. Additional background details on these methods can be found in \cref{sec:variational-autoencoder,sec:PCA-Based Baselines} in the supplementary material.
We also develop another baseline, MERL100, that represents our method trained on the original MERL dataset to demonstrate the effectiveness of our data augmentation.
% Although DeepBRDF~\cite{deepbrdf} addresses a similar task in neural material synthesis, a direct comparison is not feasible due to the unavailability of its codebase and reported metrics. \todo{DeepBRDF is cond. generation baseline. Should it be in the next sec? - Peter}

\Cref{tbl:gen-eval-metrics-all} provides a detailed comparison of baselines with our proposed metrics, while \cref{fig:uncond-spheres,fig:teaser} showcase renderings of these materials across various geometries and scenes. To achieve more complex visual effects, our synthesized materials support rendering with bump or normal maps, as well as spatially varying configurations~\cite{Jakob2022DrJit} (additional results in \cref{sec:Further Results}). The quantitative and qualitative results indicate that \ours consistently outperforms all baselines across metrics, producing diverse, high-quality, visually appealing, and perceptually realistic renderings. This demonstrates the effectiveness of \ours for neural material synthesis. Notably, materials synthesized by some baselines exhibit significant artifacts, likely due to the limitations of these simpler models in capturing the complex distribution of measured materials.
\begin{table*}
  \centering
    % \rowcolors{2}{}{LightCyan}
  \begin{tabular}{@{}rlccccccc@{}}
  \toprule
  \multicolumn{2}{c}{Metric} & Training set & PCA-A & PCA-N & VAE-A & VAE-N & MERL100 &\ours (ours) \\
  \midrule
\multicolumn{2}{c}{FID ($\downarrow$)} & 0.187 & 10.9 & 23.8 & 26.1 & 10.0 &  7.56  & \textbf{0.440} \\ \midrule
\multirow{4}{*}{MMD ($\downarrow$)}
& BRDF-L1$\times 10^{-3}$ & 2.51 & 9.05 & 9.22 & 9.09 & 5.83 &  4.30 & \textbf{4.02} \\
% & BRDF-L1-Log &  1.05 & 2.72 & 1.66 & 3.70 & 14.6 &  MERL100 & \textbf{1.21} \\
% & BRDF-NF & 0.620 & 2.35 & 1.35 & 3.13 & 12.4 & MERL100 & \textbf{0.765} \\
% & sam-brdf & \todo{} & 2.4451 & $\infty$ & 1.8517 & $\infty$ & MERL100 & \textbf{0.6651} \\
& RMSE$\times 10^2$ & 7.54 & 33.3 & 30.2 & 63.7 & 15.5  &  13.4 & \textbf{9.34} \\
& NegPSNR & $-$28.7 & $-$13.9 & $-$14.8 & $-$8.30 & $-$20.9 &   $-$22.6 & \textbf{$-$25.6} \\
& NegSSIM$\times 10$ & $-$9.55 & $-$6.74 & $-$6.29 & $-$2.68 & $-$6.86 &  $-$8.27 & \textbf{$-$9.40} \\
\midrule
\multirow{4}{*}{COV (\%) ($\uparrow$)}
& BRDF-L1 & 60.8 & 2.50 & 30 & 0.833 & 20.8 &  28.3 & \textbf{50.8} \\
% & BRDF-L1-Log & 63.3 & 8.33 & \textbf{30} & 1.67 & 0.833 & 27.5 \\
% & BRDF-NF & 64.2 & 8.33 & \textbf{30.8} & 0.833 & 0.833 & 27.5 \\
& RMSE & 55.8  & 18.3 & 28.3 & 0.833 & 16.7 &  25.0 & \textbf{50.0} \\
& NegPSNR & 56.7 & 18.3 & 28.3 & 0.833 & 18.3 &  25.0 & \textbf{50.0} \\
& NegSSIM & 59.2 & 23.3 & 16.7 & 0.833 & 17.5 &   22.5 & \textbf{51.7} \\
\midrule
\multirow{4}{*}{1-NNA (\%) ($\downarrow$)}
& BRDF-L1 & 58.8 & 100 & 95.4 & 100 & 96.7 &   92.5 & \textbf{80.0} \\
% & BRDF-L1-Log & 98.3 & 100 & \textbf{92.1} & 100 & 100 & 100 \\
% & BRDF-NF & 93.3 & 100 & \textbf{90.4} & 100 & 100 & 99.6 \\
& RMSE & 55.4 & 96.3 & 93.4 & 100 & 93.3 &   84.6 & \textbf{60.0} \\
& NegPSNR & 55.0 & 94.2 & 90.0 & 100 & 93.3 &  84.6 & \textbf{60.4} \\
& NegSSIM & 57.5 & 96.3 & 96.7 & 100 & 93.8 &   86.3 & \textbf{61.7} \\
  \bottomrule
  \end{tabular}
  \caption{Qualitative evaluation of unconditional synthesis with metrics assessing generation fidelity and diversity. 
  % $\downarrow$ indicates that a lower score is better and $\uparrow$ indicates the opposite.
  \ours significantly outperforms all baseline models across these metrics, underscoring its effectiveness in neural material synthesis.}
  \label{tbl:gen-eval-metrics-all}
\end{table*}

\subsection{Multi-Modal Conditional Synthesis}
\label{sec:conditional-synthesis}
\begin{figure}
    \centering
    \includegraphics[width=0.95\linewidth]{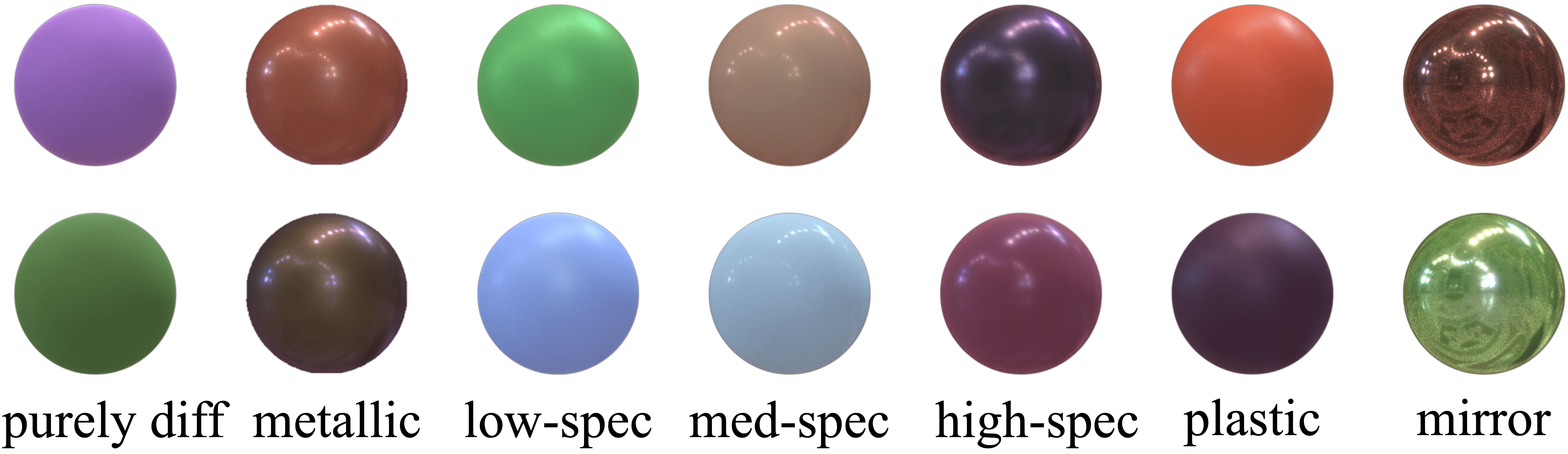}
    \caption{Synthesized materials of seven distinct categories using our novel constrained synthesis. Grounded in BRDF statistical analysis, this approach provides enhanced explainability and interpretability compared to standard conditional synthesis methods.}
    \label{fig:constrained-synthesis-vis}
\end{figure}
To further evaluate the effectiveness of our pipeline, we perform multi-modal conditional synthesis by conditioning our model on various modalities of input: material type, text description, or material images. 

For material type conditioning, we represent each of the 48 material types in the MERL dataset~\cite{Matusik2003datadriven} (\eg, \emph{acrylic}, \emph{metallic}, \emph{plastic}, \etc) using integers. The full list of material types is available in \cref{sec:Full Material Type List} in the supplementary; For text conditioning, we use descriptions derived from the MERL dataset~\cite{Matusik2003datadriven}. For additional materials in the AugMERL dataset, descriptions are assigned as follows: For RGB-permuted materials, we retain the original description but omit color-specific words (\eg, \emph{“red metallic paint”} becomes \emph{“metallic paint”}); For PCA-interpolated materials, we generate descriptions in the format \emph{“a mixture of $t_A$ and $t_B$”}, where $t_A$ and $t_B$ are the descriptions of the interpolated materials $A$ and $B$, respectively; For image-based conditioning, we use cropped single-view renderings of materials from AugMERL as input.
% to guide the synthesis based on visual references.
We encode input texts and images using CLIP encoders~\cite{radford2021learning}.

\begin{figure}[tbh]
  \centering
  \includegraphics[width=0.85\linewidth]{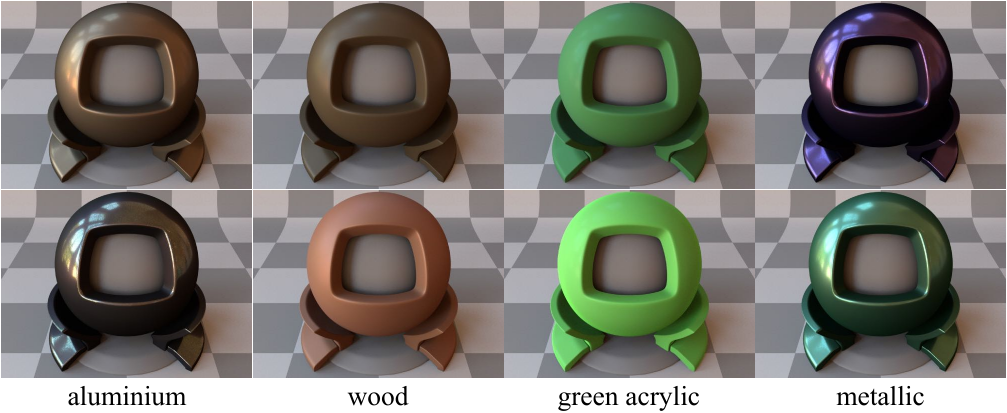}
  \caption{Type-conditioned synthesis. The synthesized materials are diverse and closely align with the input type.}
  \label{fig:type-cond}
\end{figure}

\begin{figure}[tbh]
  \centering
   \includegraphics[width=0.95\linewidth]{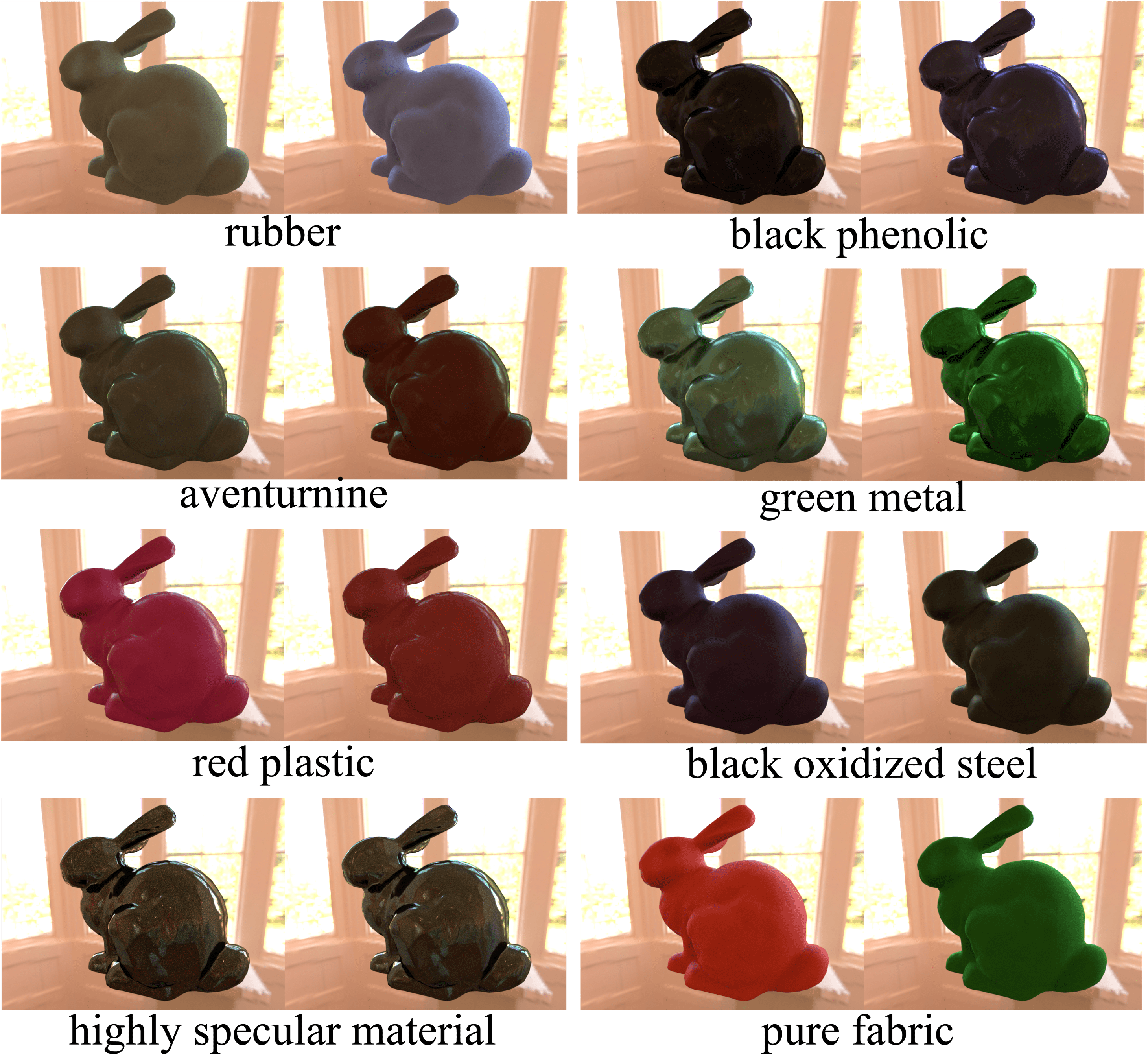}
    \caption{Text-conditioned synthesis. Synthesized materials align with the texts and generalize to unseen inputs: \emph{``green metal''}, \emph{``red plastic''}, and \emph{``highly specular material''}.}
  \label{fig:text-cond}
\end{figure}

\begin{figure}[tbh]
  \centering
  \includegraphics[width=0.95\linewidth]{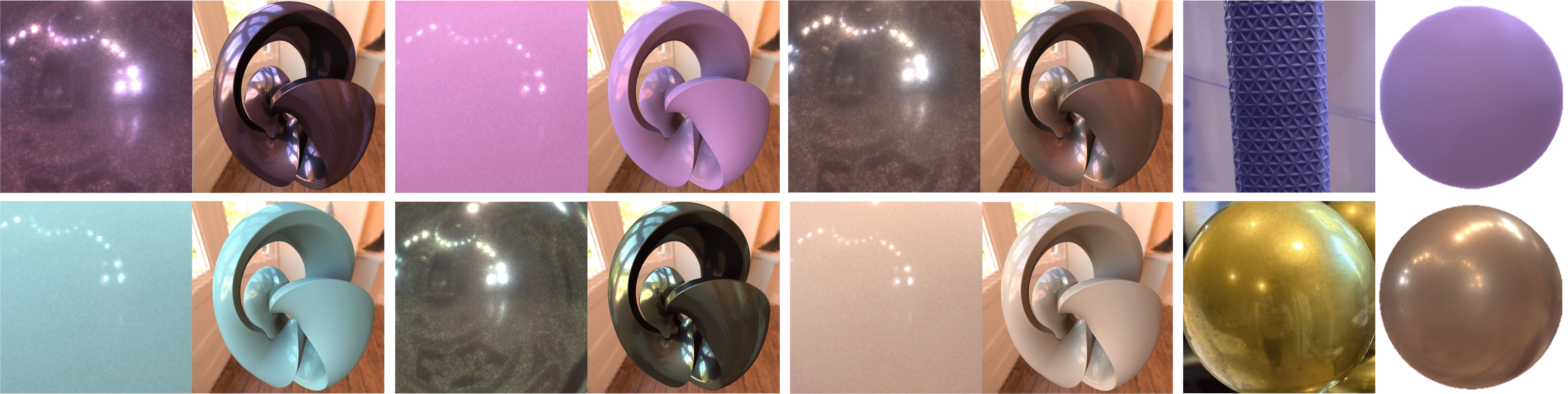}
      \caption{Image-conditioned synthesis. Each of the eight pairs consists of the input (left) and the synthesized (right). \ours effectively generates realistic materials that closely align with the conditioning images and generalizes to unseen, real-world images (last column).}
  \label{fig:image-cond}
\end{figure}

\Cref{fig:type-cond,fig:text-cond,fig:image-cond} present the results for type-, text-, and image-conditioned synthesis, respectively. Across all conditioning modes, the synthesized materials demonstrate realism, diversity, and a close alignment with the input conditions. Notably, in text- and image-conditioned synthesis, \ours effectively generalizes to unseen texts (\eg, \emph{“green metal”}, \emph{“red plastic”}, and \emph{“highly specular material”}) and real-world images, producing materials that are perceptually consistent with these previously unseen inputs.

\subsection{Constrained Synthesis}
\label{sec:Constrained Synthesis}
We classify materials into seven categories based on their reflective properties: \emph{diffuse}, \emph{metallic}, \emph{low-specular}, \emph{medium-specular}, \emph{high-specular}, \emph{plastic}, and \emph{mirror}. To enable the synthesis of materials within a specified category, we introduce a novel approach called \emph{constrained synthesis}. This statistics-based method complements our conditional pipeline by enforcing constraints on unconditionally synthesized samples, allowing for targeted material generation according to desired reflective characteristics.

We derive the theoretical upper limit for a diffuse reflectance value, \( f_\text{diffuse} \), for a constant (\ie, purely diffuse) BRDF \( f_r(\rv{\omega}_i, \rv{\omega}_o) \equiv f_\text{diffuse} \) in each color channel. For a physically valid BRDF that adheres to energy passivity~\cite{zhou2024physically}, the reflected energy must not exceed the incident energy in each channel. Thus, we have:
\begin{align}
    1 &\geq \int_{H^2}  f_r(\rv{\omega}_i, \rv{\omega}_o) \cos \theta_o \, \mathrm{d} \rv{\omega}_o \\
    &= f_\text{diffuse} \int_{H^2} \cos \theta_o \, \mathrm{d} \rv{\omega}_o = \pi f_\text{diffuse}
    \Rightarrow f_\text{diffuse} \leq \frac{1}{\pi}.
\end{align} 

Building on this observation and a statistical analysis of the MERL's mean and maximum reflectance values across color channels and material types (\cref{sec:MERL Statistical Analysis Details} in the supplementary), we propose a set of rules for categorizing materials into seven types. These rules enable the selection of synthesized materials based on desired material characteristics. Unlike many black-box machine learning approaches, this method is rooted in BRDF analysis, offering inherent explainability and interpretability. Below, we outline two of these rules, with the full set detailed in \cref{sec:Full Set of Constrained Synthesis Rules}.
\begin{itemize}
\item Purely diffuse materials: The reflectance values in all directions do not exceed the diffuse threshold \( f_\text{diffuse} \), allowing for only \( e \coloneqq 8 \times 10^4 \) exceptions:
\begin{equation}
    \lvert \{ (\rv{\omega}_o, \rv{\omega}_i) \in \mathbb{R}^6 \mid \left\| f_r(\rv{\omega}_o, \rv{\omega}_i) \right\|_\infty > f_\text{diffuse} \} \rvert < e,
\end{equation}
where \( \left\| \cdot \right\|_\infty \) denotes the maximum reflectance value among the three color channels.

\item Metallic materials: The reflectance values in all directions exceed the diffuse threshold \( f_\text{diffuse} \):
\begin{equation}
    \forall (\rv{\omega}_o, \rv{\omega}_i) \in \mathbb{R}^6, \left\| f_r(\rv{\omega}_o, \rv{\omega}_i) \right\|_\infty > f_\text{diffuse}
\end{equation}
\end{itemize}

\Cref{fig:constrained-synthesis-vis} shows materials generated through our constrained synthesis, incorporating the proposed filtering rules to ensure that the synthesized outputs match the specified material categories. The results demonstrate that the synthesized materials effectively exhibit the characteristics of the desired categories.

\subsection{Ablation Study}
\label{sec:ablation-study}
To assess the impact of our augmented material dataset AugMERL, we further train our model on the original MERL dataset in the unconditional synthesis task. We report both quantitative and qualitative results for this model, presented in the ``MERL100'' columns of \cref{tbl:gen-eval-metrics-all} and \cref{fig:uncond-spheres}, respectively. The results indicate that the model trained on AugMERL exhibits higher quality and diversity compared to the one trained on MERL, demonstrating the effectiveness of our augmented dataset in enhancing the synthesis pipeline.

Additionally, we conduct sparse BRDF reconstruction and BRDF compression experiments following a previous method~\cite{gokbudak2023hypernetworks}. For sparse reconstruction, we set the sample size to \( N=4000 \), while for compression, we use a latent dimension of \( 40 \). In both experiments, we train the model on either the original MERL dataset or the AugMERL dataset. The results, summarized in \cref{tab:augmerlablation}, demonstrate that training on AugMERL consistently enhances material quality across all evaluated metrics, further validating the effectiveness of our augmented dataset.
\begin{table}
\centering
{
\setlength\tabcolsep{3.5pt}
\begin{tabular}{@{}rcc|cc@{}}
\toprule
\multicolumn{1}{c}{\multirow{2}{*}{Metric}} & \multicolumn{2}{c|}{Sparse reconstruction} & \multicolumn{2}{c}{Compression} \\
\cmidrule(lr){2-5}
 & MERL & AugMERL & MERL & AugMERL \\
\midrule
 PSNR ($\uparrow$) & 32.2 & \textbf{36.3} & 45.2 & \textbf{48.3} \\
 Delta E ($\downarrow$) & 2.1 & \textbf{1.8} & 0.693 & \textbf{0.623} \\
 SSIM ($\uparrow$) & 0.972 & \textbf{0.983} & \textbf{0.994} & \textbf{0.994} \\
 \bottomrule
\end{tabular}
}
\caption{Quantitative comparison of training on MERL versus AugMERL in the sparse BRDF reconstruction and BRDF compression experiments. The results demonstrate that training on AugMERL consistently enhances performance across all metrics.}
\label{tab:augmerlablation}
\end{table}
\section{Conclusion and Future Work}
\label{sec:conclusion}
% High-quality material synthesis is essential for realistic renderings. 
We introduced \ours, a \textbf{m}ulti-\textbf{m}odal \textbf{m}aterial \textbf{s}ynthesis approach with \textbf{h}yperdiffusion. Using neural fields as the core representation, we trained hyperdiffusion on their weights, demonstrating its ability to generate high-quality, diverse materials. Additionally, we contribute two material datasets and three BRDF metrics for future research. At this stage, our method does not account for physical correctness. Promising directions for future work include developing physically accurate neural representations of BRDFs and extending the approach to support more complex materials.
% Nevertheless, our method is currently limited to uniform materials. Future work will explore more complex cases, such as spatially varying BRDFs.
% High-quality material synthesis is critical for achieving realistic, convincing renderings. In this work, we introduced \ours, an innovative \textbf{m}ulti-\textbf{m}odal \textbf{m}aterial \textbf{s}ynthesis with \textbf{hy}perdiffusion. Using neural fields as the core representation for materials, we trained our hyperdiffusion on their weights, demonstrating its effectiveness in generating high-quality, diverse materials through extensive experiments. Additionally, we contribute two material datasets and three BRDF distributional metrics to facilitate future research.

% However, our current method is limited to uniform materials. Extending to accommodate more complex materials, such as spatially varying BRDFs, may require additional mechanisms, which we aim to explore in future work.

\clearpage
{\small
\bibliography{main}
}

% WARNING: do not forget to delete the supplementary pages from your submission
\clearpage
\appendix
% \maketitlesupplementary

\section{Related Work on Material Acquisition and Databases}
\label{sec:Related Work on Material Acquisition and Databases}
Capturing real-world material appearance requires lengthy acquisition times and substantial storage. Traditional capture uses four-axis gonioreflectometers~\cite{gonio, White98} with later advancements enhancing angular coverage, wavelength resolution, efficiency, and accuracy~\cite{White98}. Systems range from basic sensor setups, like camera arrays~\cite{weinmann2015advances}, varied illumination sources~\cite{haindl2013visual}, and specific geometries~\cite{marschner99, kaleidoscop, weinmann2015material},
to advanced configurations for capturing full material patches~\cite{danabtf, weinmann-2014}. Filip and Vávra contributed a database of 150 materials, many exhibiting anisotropic properties~\cite{filip2014template}. These material databases are continuously being augmented by later methods~\cite{ngan2005experimental, dupuy2018adaptive}. These improvements have enabled the creation of extensive BRDF material databases~\cite{Matusik2003datadriven, rgl2018, NielsenPCA2015}, which underpin the datasets in our work.

\section{Additional Background}
\subsection{Principal Component Analysis}
\label{sec:principal-component-analysis}
Principal component analysis (PCA)~\cite{abdi2010principal}, also known as Karhunen-Loève transform or Hotelling transform, is a linear dimensionality reduction technique. We utilize PCA for data augmentation in \cref{sec:principal-component-analysis} and baseline models in \cref{sec:unconditional-synthesis}.

Given a dataset of $n$ samples in a high-dimensional space $\mathcal{D} \in \mathbb{R}^{n \times d}$, PCA seeks to linearly transform the data onto a lower-dimensional space $\mathbb{R}^k$ spanned by $k$ \emph{principal components} capturing the primary variance of the data. In this reduced space, synthetic data points are generated by sampling from a Gaussian distribution with the same mean and variance as the original data in each principal component direction. These newly sampled points are then mapped back to the original feature space using the inverse PCA transformation. This method allows the creation of new data samples that maintain the underlying structure and variance characteristics of the original dataset, which can be useful for data augmentation and improving model robustness.

\subsection{Diffusion Model}
\label{sec:more-diffusion-model}
The forward and backward processes in hyperdiffusion are modeled as Markov chains with a total timestep \( T \) and learnable parameter \( \rv{\eta} \):
\begin{align}
&q(\rv{x}_1, \ldots, \rv{x}_T \vert \rv{x}_0) = \prod_{t=1}^T q(\rv{x}_t \vert \rv{x}_{t-1}), \\
&p_\rv{\eta}(\rv{x}_0, \ldots, \rv{x}_T) = p_\rv{\eta}(\rv{x}_t) \prod_{t=1}^T p_\rv{\eta}(\rv{x}_T \vert \rv{x}_{t-1}).
\end{align}
In the forward process, starting with the original data \( \rv{x} = \rv{x}_0 \), we iteratively add Gaussian noise at each step:
\begin{equation}
    q(\rv{x}_t \vert \rv{x}_{t-1}) := \mathcal{N}(\rv{x}_t; \sqrt{1-\beta_t} \rv{x}_{t-1}, \beta_t I),
\end{equation}
where \(\{\beta_t\}_{t=1}^T\) defines the variance schedule. Each noisy vector, paired with the sinusoidal embedding of the timestep, is passed through a linear projection layer. The output projections are then combined with a learnable positional encoding vector. As the forward process progresses, \( p(\rv{x}_T) \) converges towards a standard Gaussian distribution, \( \mathcal{N}(0, I) \).

In the backward process, the transformer takes these inputs and produces denoised tokens, which are passed through a final output projection layer to generate the predicted noise. To train a learnable model \(\rv{\epsilon}_\rv{\eta}(\rv{x}_t, t)\) parameterized by \(\rv{\eta}\), we minimize the score matching objective:
\begin{equation}
    \mathcal{L}_\text{HD}(\rv{\eta}) := \mathbb{E}_{\rv{x}_0,t \sim \mathcal{U}(1,T), \rv{\epsilon} \sim \mathcal{N}(0, I)} \left[ \| \rv{\epsilon} - \rv{\epsilon}_\rv{\eta}(\rv{x}_t,t) \|_2^2 \right],
\end{equation}
where \(\mathcal{U}(1,T)\) represents the uniform distribution over \(\{1,2,\ldots,T\}\). This objective encourages the model to accurately predict the noise \(\rv{\epsilon}\), effectively guiding the denoising process.

During inference, the network enables sampling via an iterative process~\cite{song2022ddim}, leveraging the factorization of the learned distribution as
\begin{equation}
\label{eq:ddpm-gen-supp}
    p_\rv{\eta}(\rv{x}) = p(\rv{x}_T) p_\rv{\eta}(\rv{x}_0 \vert \rv{x}_T) = p(\rv{x}_T) \prod_{t=1}^T p_\rv{\eta} (\rv{x}_{t-1} \vert \rv{x}_t)
\end{equation}
for $p(\rv{x}_T) := \mathcal{N}(0, I)$. For conditional sampling, we employ classifier-free guidance (CFG)~\cite{ho2022cfg} (please refer to \cref{sec:Conditional Sampling with Classifier-Free Guidance} for the algorithm).

\subsection{Attention Mechanism}
\label{sec:attention-module}
The attention mechanism~\cite{vaswani2017attention} allows models to focus on specific parts of input data, dynamically assigning different levels of ``attention'' or importance to different elements. The attention mechanism enables models to learn which parts of the input sequence are most relevant to predicting each output token. This process improves performance by allowing models to prioritize relevant information and ignore irrelevant details, especially in long sequences.

In modern applications, attention modules are widely used in natural language processing (NLP), computer vision, and beyond. The transformer architecture~\cite{vaswani2017attention}, which relies heavily on the self-attention mechanism, has become foundational in NLP models, including BERT~\cite{devlin2018bert} and GPT~\cite{mann2020language}. Attention allows these models to capture complex dependencies between words in a sentence, regardless of their distance from each other, leading to significant advancements in tasks like language translation, sentiment analysis, and image processing.

The attention mechanism computes a weighted combination of values based on the relevance of each value to a given query. In the context of self-attention (or scaled dot-product attention) in transformer models~\cite{vaswani2017attention}, this process involves three main components: queries ($\rv{q}$), keys ($\rv{k}$), and values ($\rv{v}$). In summary, the attention mechanism dynamically focuses on relevant parts of the input by computing similarity scores between queries and keys, normalizing these scores, and using them to weight the values. This allows models to capture dependencies across elements in a sequence, making attention a powerful tool for handling long-range dependencies in data.

Given an input sequence of embeddings $\rv{X}$ (\eg, word embeddings in NLP or patch embeddings in vision), we first transform it into three different linear projections:
\begin{align}
    \rv{q} &= \rv{X}W_\rv{q};\\
    \rv{k} &= \rv{X}W_\rv{k};\\
    \rv{v} &= \rv{X}W_\rv{v},
\end{align}
where \( W_\rv{q} \), \( W_\rv{k} \), and \( W_\rv{v} \) are learnable weight matrices. These projections represent the queries, keys, and values, respectively. The attention score between each query and key is then computed as the dot product \( \rv{q} \cdot \rv{t} \). This results in a matrix of scores that represents the similarity between each element in the sequence. To stabilize gradients and prevent large values in the dot-product computation, the scores are scaled by the square root of the dimensionality of the queries/keys \( \sqrt{d_\rv{k}} \). The scaled scores are \( \frac{\rv{q} \cdot \rv{t}}{\sqrt{d_k}} \).

The scaled scores are passed through a softmax function, producing attention weights that sum to 1. This step converts the scores into probabilities, indicating the relevance of each value with respect to each query. These attention weights are used to compute a weighted sum of the values. Specifically, the output of the attention mechanism is
\begin{equation}
f_\text{attention}(\rv{q}, \rv{k}, \rv{v}) = f_\text{softmax}\left(\frac{\rv{q} \cdot \rv{k}}{\sqrt{d_k}}\right) \rv{v}.
\end{equation}
This produces a context vector for each query that incorporates information from all other elements in the sequence, weighted by their relevance.

In practice, multiple attention heads are used in parallel. Each head learns different aspects of the input by using different projections \( W_\rv{q}, W_\rv{k}, \) and \( W_\rv{v} \). The outputs from each head are then concatenated and linearly transformed to produce the final output.

\subsection{Variational Autoencoder}
\label{sec:variational-autoencoder}
We adopt variational autoencoders (VAEs)~\cite{kingma2013auto} as one of the baseline models in \cref{sec:unconditional-synthesis}. VAEs are probabilistic generative models designed to capture the underlying probability distribution of a given dataset $\mathcal{D}$.

A VAE consists of a parameterized encoder $\vaeenc$ and decoder $\vaedec$, defined by parameters $\rv{\psi}$ and $\rv{\zeta}$, respectively. Assuming that the latent variables $\rv{z} \in \mathbb{R}^{D_\rv{z}}$ follow a prior distribution $\vaeprior$, both the encoder and decoder are jointly optimized to maximize the \emph{evidence lower bound (ELBO)} on the likelihood of the data:
\begin{equation}
\begin{aligned}
    \mathcal{L}_\text{ELBO} (\rv{\psi}, \rv{\zeta}; \rv{x}) := \, &\mathbb{E}{\vaeenc} \left[ \log \vaedec \right] \\ &- \mathcal{D}_\text{KL} \left( \vaeenc, \vaeprior \right),
\end{aligned}
\end{equation}
where $\mathcal{D}_\text{KL}$ represents the Kullback-Leibler divergence between the approximate posterior and the prior distribution~\cite{csiszar1975divergence}. This probabilistic framework enables VAEs to effectively approximate the true data distribution, facilitating robust generative modeling of complex data.

\subsection{K-Means Clustering}
\label{sec:k-means-clustering}
We adopt K-means clustering in constrained synthesis (\cref{sec:Constrained Synthesis}), where unconditional materials are filtered out by statistical information rather than a neural network. K-means clustering is a commonly used unsupervised method to partition \(n\) data samples \(\rv{X} \in \mathbb{R}^{n\times p}\) into \(k\)
clusters \(\mathcal{S}=\{S_1, ..., S_{k}\}\), minimizing the distance between each sample and its center
\begin{equation}
\mu_i = \frac{1}{|S_i|} \sum_{\mathbf{x_i}\in S_i}\mathbf{x_i}.
\end{equation}
The optimal partition can be computed via the following objective 
\begin{equation}
\mathcal{S}^* = \arg \min_{\mathcal{S}} \sum_{i=0}^{k-1}\sum_{\mathbf{x_i}\in S_i} ||\mathbf{x_i} -  \mu_i||^2.
\end{equation}
Given the assigned clusters, we can obtain the
classification decision boundary directly.

\section{Model Implementation Details}
\label{sec:Model Details}

\subsection{Neural Field}
The dimensionality of the flattened weights for our neural field $f_r^\rv{\xi}$ is  $D_\text{NF}=675$.

\subsection{Transformer Backbone in Hyperdiffusion}
\label{sec:Transformer Backbone in Hyperdiffusion}
The input and output tokens are mapped to vectors with learnable embedders. Sinusoidal positional encoding is also used. The feed-forward networks are single layer MLPs with ReLU activation functions.

The encoder network contains multiple identical layers, each with a feed-forward sublayer after multi-head attention. A residual connection is employed for each sublayer, followed by the layer normalization. The decoder is similar, but with an extra per-layer multi-head attention at the end, receiving the the encoder output.

\begin{figure}
  \centering
  \includegraphics[width=0.96\linewidth]{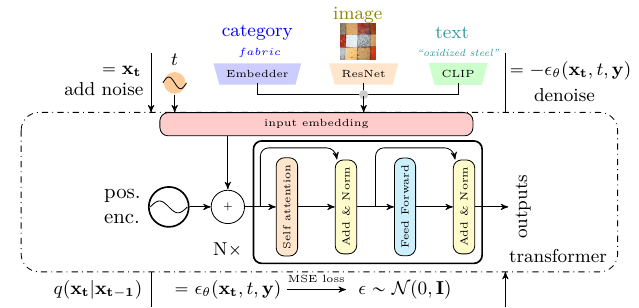}
  \caption{Transformer as hyperdiffusion backbone.}
  \label{fig:transformer_model}
\end{figure}

\cref{fig:transformer_model} illustrates the transformer architecture.

\subsection{PCA-Based Baselines}
\label{sec:PCA-Based Baselines}
For PCA-based baselines introduced in the unconditional synthesis in \cref{sec:unconditional-synthesis}, the dimensionalities of the reduced spaces for PCA-A and PCA-N are 300 and 100, respectively.

\subsection{VAE-Based Baselines}
In this section, we detail the model hyperparameters for VAE-based baselines introduced in the unconditional synthesis in \cref{sec:unconditional-synthesis}.

For VAE-A, to reduce the input complexity, we downsample it from $90 \times 90 \times 180$ to $45 \times 45 \times 90$ and upsample the synthetic results
using nearest neighbor algorithm. The hidden dimension is 256 and the latent space dimension is 300. 
% The model is trained with batch size 16, learning rate $5 \times 10^{-3}$ until convergence. 

For VAE-N, the VAE architecture contains MLP-based encoder and decoder, each with 4 layers. The input dimension is $D_\text{NF}$, and the hidden and the latent space dimensions are both 300. The likelihood of the reconstructed data is measured by the mean squared error (MSE).
% The model is trained with batch size 16, learning rate $10^{-3}$ until convergence.

\section{Experiment Details}
\label{sec:Experiment Details}
\subsection{Training Details}
\label{sec:Training Details}
In the neural field fitting, we set the batch size to 512, the number of epochs to 100, and the learning rate to $5\times 10^{-3}$.

When training the hyperdiffusion model, we set the total timestep $T$ to 100, the batch size to 512, the number of epochs to 700 for unconditional synthesis and an additional 200 for conditional synthesis, and the learning rate to adaptive from $5\times10^{-4}$ to $5\times10^{-6}$. The experiments are run on an NVIDIA GeForce RTX 4090 GPU.

\subsection{Conditional Sampling with Classifier-Free Guidance}
\label{sec:Conditional Sampling with Classifier-Free Guidance}
We employ classifier-free guidance (CFG)~\cite{ho2022cfg} for conditional sampling on the hyperdiffusion. We present the algorithm in \cref{alg:ddim}. Notice that we require $\omega > -1$ for conditional synthesis. Otherwise, the model downgrades to unconditional synthesis when $\omega=-1$.
\begin{algorithm}
  \caption{Conditional sampling with classifier-free guidance (CFG)}\label{alg:ddim}
  \begin{algorithmic}
    \Require total timestep $T$
  \Require guidance scale $\omega \geq -1$
    \Require conditional context $\rv{y}$
  \Require variance schedule $\{\beta_t\}_{t=1}^T$
  \State  $\rv{x}_T \sim \mathcal{N}(0, I_{D_\text{NF}})$ \Comment{sample $\rv{x}_T$ from prior}
  \For{$t=T$ to $1$}
  \State $\rv{\epsilon}_\rv{\eta}^\text{CFG}(\rv{x}_t, \rv{y}, t) = (1+\omega)\rv{\epsilon_\eta}(\rv{x}_t, \rv{y}, t) - \omega \rv{\epsilon_\eta}(\rv{x}_t, \emptyset, t)$
  \State $\alpha_t = \prod_{i=1}^t \sqrt{1-\beta_i}$
  \State $\gamma_t = \prod_{t=1}^{t} \alpha_t$
  \State $\rv{x}_t^\text{CFG} = \frac{1}{\sqrt{\gamma_{t}}}(\rv{x}_t - \sqrt{1- \gamma_{t}} \rv{\epsilon_\eta^\text{CFG}}(\rv{x}_t, \rv{y}, t))$ 
  \State $\rv{x}_{t-1}= \sqrt{\gamma_{t-1}} \rv{x}_t^\text{CFG} + \sqrt{1-\gamma_{t-1}} { \rv{\epsilon_\eta^\text{CFG}}(\rv{x}_t, \rv{y}, t)}$
  \EndFor
  \State \Return $\rv{x}_0$  
  \end{algorithmic}
\end{algorithm}

\subsection{Full Material Type List}
\label{sec:Full Material Type List}
We include the full list of 48 material types used in the type-conditional synthesis in \cref{sec:conditional-synthesis}: \emph{acrylic}, \emph{alum-bronze}, \emph{alumina-oxide}, \emph{aluminium}, \emph{aventurnine}, \emph{brass}, \emph{wood}, \emph{chrome-steel}, \emph{chrome}, \emph{colonial-maple}, \emph{color-changing-paint}, \emph{delrin}, \emph{diffuse-ball}, \emph{fabric}, \emph{felt}, \emph{fruitwood}, \emph{grease-covered-steel}, \emph{hematite}, \emph{ipswich-pine}, \emph{jasper}, \emph{latex}, \emph{marble}, \emph{metallic-paint}, \emph{natural}, \emph{neoprene-rubber}, \emph{nickel}, \emph{nylon}, \emph{obsidian}, \emph{oxidized-steel}, \emph{paint}, \emph{phenolic}, \emph{pickled-oak}, \emph{plastic}, \emph{polyethylene}, \emph{polyurethane-foam}, \emph{pvc}, \emph{rubber}, \emph{silicon-nitrade}, \emph{soft-plastic}, \emph{special-walnut}, \emph{specular-fabric}, \emph{specular-phenolic}, \emph{specular-plastic}, \emph{stainless-steel}, \emph{steel}, \emph{teflon}, \emph{tungsten-carbide}, \emph{two-layer}.

\subsection{Full Set of Constrained Synthesis Rules}
\label{sec:Full Set of Constrained Synthesis Rules}
We include the full set of seven rules used in constrained synthesis:
\begin{itemize}
\item Purely diffuse materials: the reflectance values from all directions do not exceed the diffuse threshold $f_\text{diffuse}$, with only $e:=8\times10^4$ exceptions:
\begin{equation}
    \lvert \{ (\rv{\omega}_o, \rv{\omega}_i) \in \mathbb{R}^6 \mid \left\| f_r(\rv{\omega}_o, \rv{\omega}_i) \right\|_\infty > f_\text{diffuse} \} \rvert < e,
\end{equation}
where $\left\| \cdot \right\|$ is the $\ell^\infty$-norm that selects the maximum component from the three color channels of reflectance values.

\item Metallic materials: the reflectance values from all directions exceeds the diffuse threshold $f_\text{diffuse}$:
\begin{equation}
    \forall (\rv{\omega}_o, \rv{\omega}_i) \in \mathbb{R}^6, \left\| f_r(\rv{\omega}_o, \rv{\omega}_i) \right\|_\infty > f_\text{diffuse}
\end{equation}

\item Specular materials: Through our statistical analysis, we identify two specular thresholds $f_\text{specular}^{(1)}:=100$ and $f_\text{specular}^{(2)}:=600$. The materials can be classified as low\mbox{-}, mid-, or high-specular if the maximum reflectance value fall in the range $[f_\text{diffuse}, f_\text{specular}^{(1)}), [f_\text{specular}^{(1)}, f_\text{specular}^{(2)})$, and $[f_\text{specular}^{(2)}, \infty)$, respectively.
  
\item Plastic materials: materials whose specular part is white. In other words, if the reflectance value exceeds $f_\text{diffuse}$ for some direction, the difference between any two color channels should be smaller than a tolerance $\delta_\text{plastic}$, which we set to be 5\% of the maximum reflectance value.
  
\item Mirror-like materials: the specular lobe in the polar diagram is narrow. In order to quantify this, we observe that under the Rusinkiewicz reparametrization (\cref{sec:NBRDF Fitting}), the evaluation at $f_r(\theta_{\rv{h}}=0, \theta_{\rv{d}}=0, \varphi_{\rv{d}}=0)$ is the peak of the lobe. Incrementing $\theta_\rv{h}$ decreases the BRDF. The width $w$ of the lobe can then be defined as the value of $\theta_\rv{h}$ for which the reflectance value drops to half of the the peak $\frac{f_r(0, 0, 0)}{2}$. Empirically, we found that if the width $w < w_\text{mirror}:=0.349$, the material exhibits mirror-like behavior.
\end{itemize}

\section{Metric Details}
\label{sec:Metric Details}
\subsection{Rendering-Based Metrics}
\label{sec:Rendering-Based Metrics}
Given two rendered images
\(I_1, I_2\colon \mathbb{R}^w \times \mathbb{R}^h \rightarrow [0,1]^c\), where $w,h$ are the width and height of the images, respectively, and $C$ is the number of channels, we propose the following rendering-based metrics assessing the similarity and reconstruction quality between the two images.

\paragraph{Root mean squared error (RMSE)} RMSE checks if pixel values at the same coordinates match.
\[
\mathcal{L}_\text{RMSE}(I_1,I_2) := \sqrt{\frac{1}{wh} {\sum_{i=1}^{w} \sum_{j=1}^{h} \left( I_1(i,j)-I_2(i,j) \right)^2. } }
\]
Note that RMSE depends strongly on the image intensity scaling. RMSE aims for lower value for better performance.

\paragraph{Peak signal-to-noise ratio (PSNR)} is the scaled mean squared error (MSE). Given the
maximum pixel value \(p\), \ie, peak signal, PSNR is defined as
\[
\mathcal{L}_\text{PSNR}(I_1, I_2) =  10 \log_{10} \frac{p^2}{\mathcal{L}_\text{RMSE}^2(I_1, I_2)}.
\]
PSNR measures the image reconstruction quality and aims for higher values.

\paragraph{Structural similarity index measure (SSIM)} SSIM~\cite{SSIM04} is a perception-based metric that measures the perceptual similarity of the two images. The computation of SSIM is based on three comparison measurements between the two images: luminance ($l$), contrast ($c$), and structure ($s$) defined as
\begin{align}
   & l(I_1, I_2) := \frac{2 \mu_1 \mu_2 + c_1}{\mu_1^2 + \mu_2^2 + c_1};\\
   & c(I_1, I_2) := \frac{2 \sigma_{I_1} \sigma_{I_2} + c_2}{\sigma_{I_1}^2 + \sigma_{I_2}^2 +c_2 };\\
   & s(I_1, I_2) := \frac{\sigma_{I_1 I_2} + c_3}{\sigma_{I_1} \sigma_{I_2} + c_3},
\end{align}
respectively, where $\mu_I, \sigma_I$, and $\sigma_{I_1I_2}$ are the mean, standard deviation, and variance of the images:
\begin{align}
    & \mu_I :=  \frac{1}{w \cdot h} {\sum_{i=1}^{w} \sum_{j=1}^{h}  I(i,j) };\\
    & \sigma_I := \sqrt{ \frac{1}{w \cdot h- 1} {\sum_{i=1}^{w} \sum_{j=1}^{h}  \left( I(i,j)-\mu_I \right)^2 }};\\
    & \sigma_{I_1I_2} := \frac{1}{w \cdot h - 1} \sum_{i=1}^{w} \sum_{j=1}^h \left(I_1(i,j)-\mu_{I_1})(I_2(i,j)-\mu_{I_2} \right)
\end{align}
and 
\begin{align}
    & c_1 := (k_1 L)^2; \\
    & c_2 := (k_2 L)^2; \\
    & c_3 := \frac{c_2}{2}
\end{align}
are the variables to stabilize the division with weak denominator where $k_1$ and $k_2$ are coefficients defaulted to 0.01 and 0.03, respectively, and $L$ is the dynamic range of the pixel-values typically chosen to be $2^l-1$ where $l$ is the number of bits per pixel.

SSIM is then defined as a weighted combination of the above comparative measures with exponential weights $a,b,c > 0$:
\begin{equation}
\mathcal{L}_\text{SSIM}(I_1, I_2) := l^a(I_1, I_2) c^b(I_1, I_2) s^c(I_1, I_2).
\end{equation}

SSIM aims for higher values for better performance.

\subsection{Distributional Metrics Validation}
\label{sec:proposed-metrics-validation}
In this section, we explore the effectiveness of the proposed novel material distributional metrics (\cref{sec:performance-metrics}) by examining the nearest neighbor materials under certain distances.

\begin{figure*}
  \centering
  \begin{subfigure}{0.48\linewidth}
      \includegraphics[width=\linewidth]{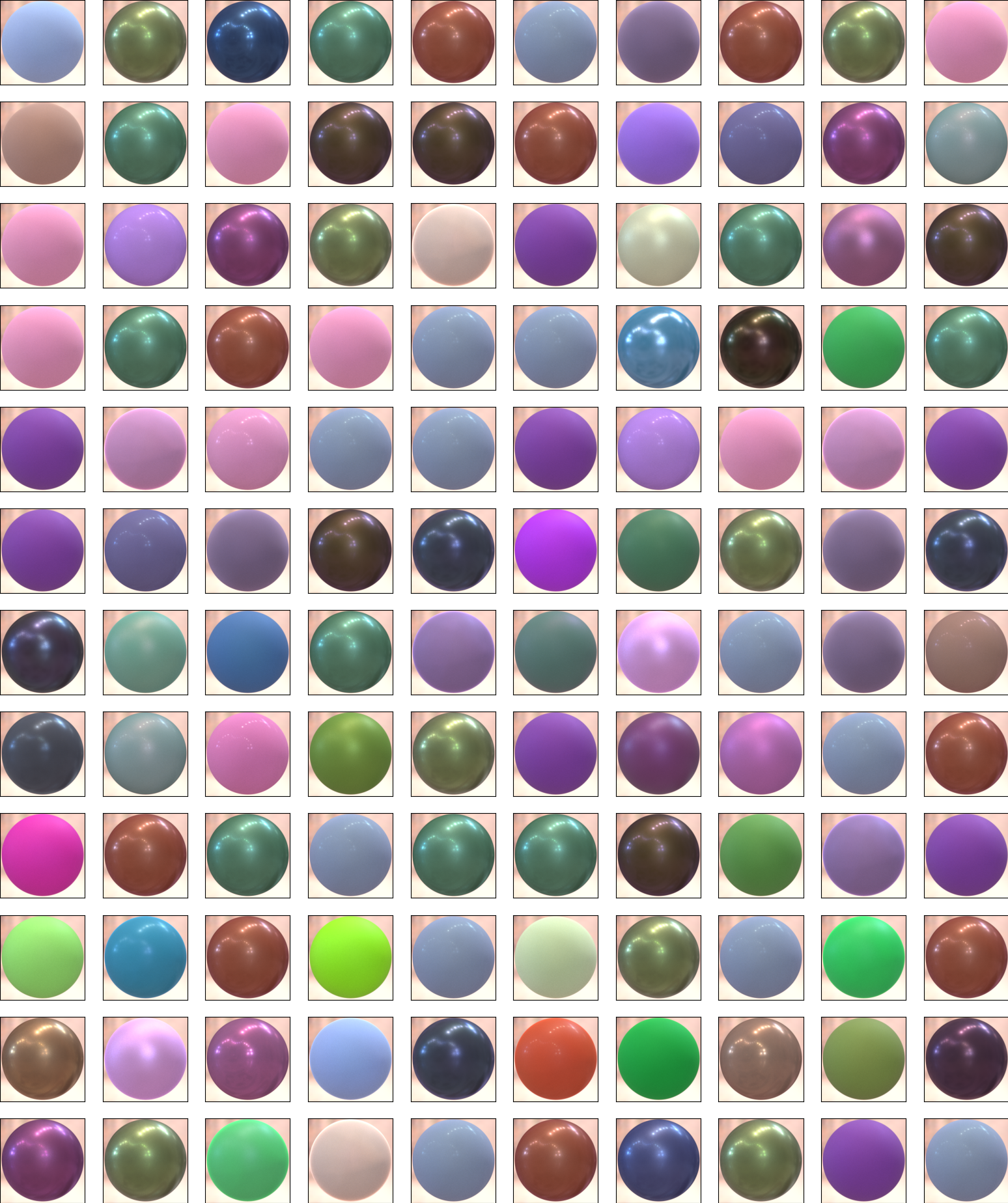}
      \caption{Synthetic set}
  \end{subfigure}
  \hfill
  \begin{subfigure}{0.48\linewidth}
      \includegraphics[width=\linewidth]{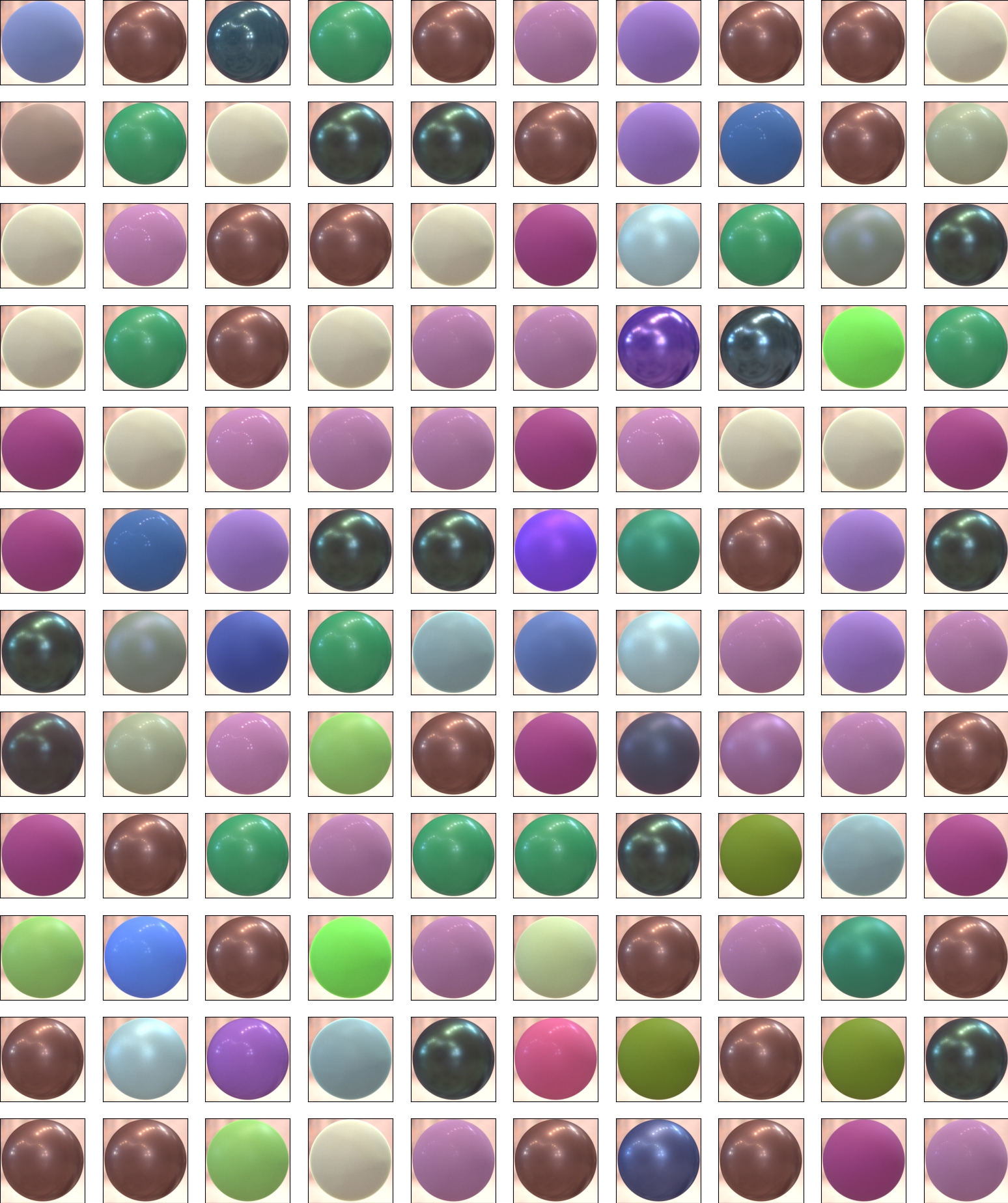}
      \caption{Nearest neighbor reference}
  \end{subfigure}

  \caption{Nearest reference under $d_\text{BRDF-L1-Log}$ (\cref{eq:brdf-distance-l1-log}): for each synthetic material in (a), we identify its nearest neighbor in the reference set under $d_\text{BRDF-L1-Log}$ presented at the same grid position in (b). The nearest neighbor metric is capable of matching most BRDFs in terms of reflective behaviors.}
  \label{fig:eval-uncond-metrics}
\end{figure*}

In \cref{fig:eval-uncond-metrics}, we demonstrate the nearest reference BRDF to each synthetic sample under the distance
\begin{equation}
\label{eq:brdf-distance-l1-log} 
    d_\text{BRDF-L1-Log} := \underset{\theta_{\rv{h}}, \theta_{\rv{d}}, \varphi_{\rv{d}}}{\mathbb{E}} \Big[ \left| \log \left(1 + f_r \right) - \log \left(1 + f_r' \right) \right| \Big],
\end{equation}
which is used to compute the distributional metrics MMD, COV, and 1-NNA. While there exist mismatches due to the limited size of the reference set, we see that proposed BRDF distance is capable of matching most BRDFs in terms of reflective behaviors, leading to plausible BRDF distributional metrics.

\begin{figure*}
  \centering
  \begin{subfigure}{0.48\linewidth}
    \includegraphics[width=\linewidth]{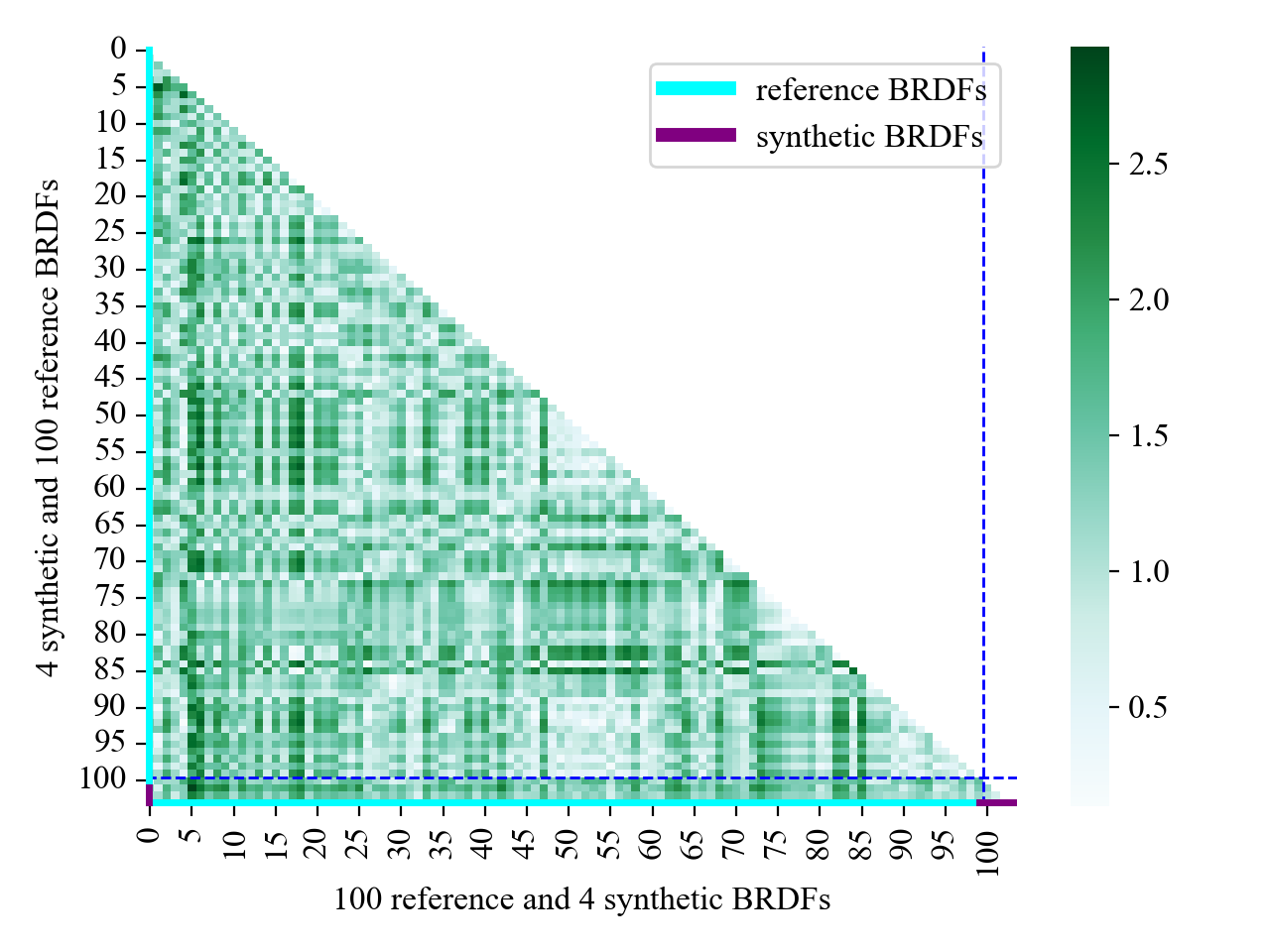}
      \caption{Pairwise distance of synthetic and all reference BRDFs}
  \end{subfigure}
  \begin{subfigure}{0.48\linewidth}
    \includegraphics[width=\linewidth]{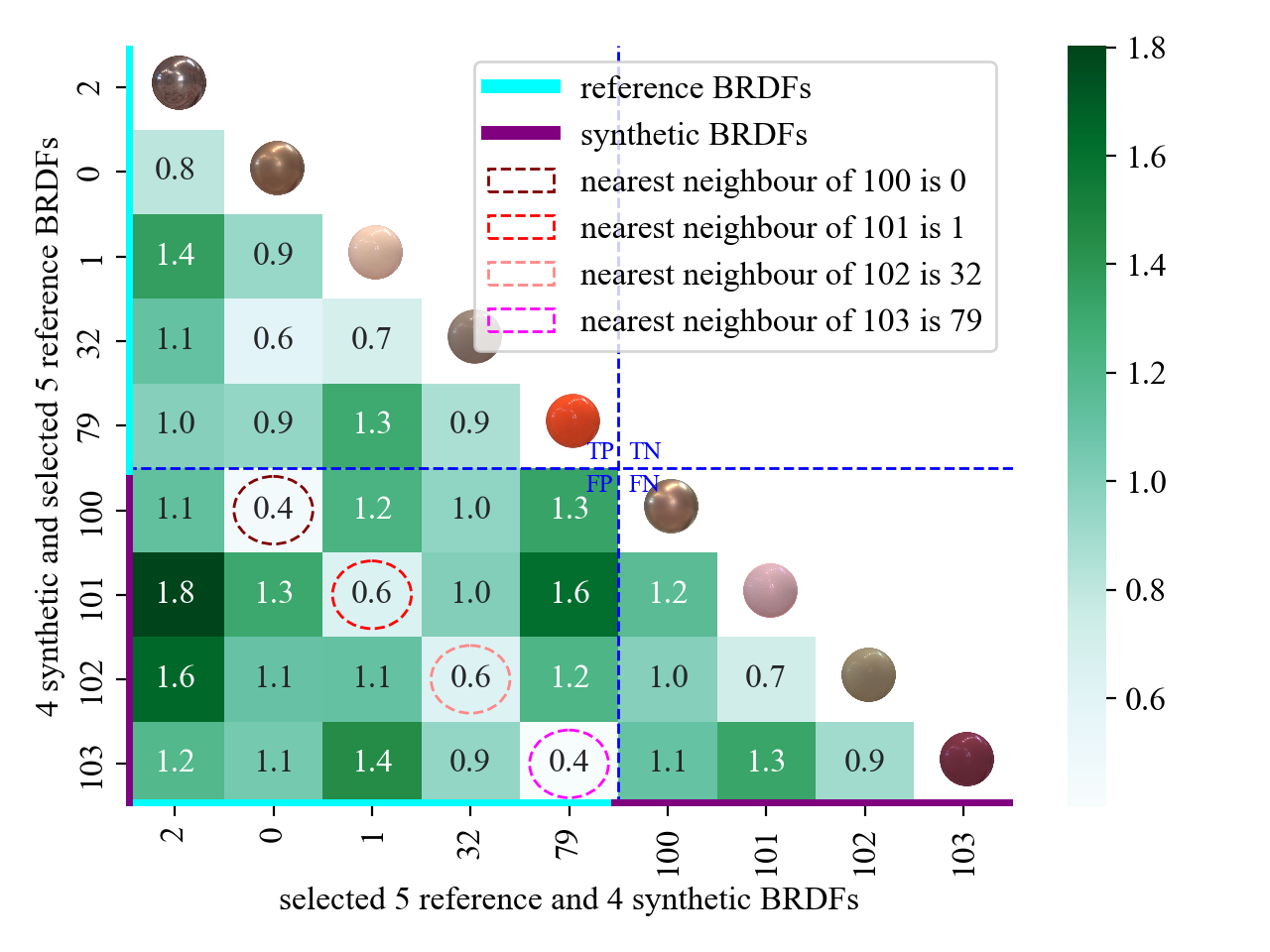}
      \caption{Pairwise distance of synthetic and selected reference BRDFs}
  \end{subfigure}
  \caption{Nearest neighbor information illustrated by pairwise mean squared logarithmic distance. We can see that the nearest neighbor is captured effectively with the underlying distance.}
  \label{fig:confusion_matrix}
\end{figure*}

In \cref{fig:confusion_matrix}, we further visualize the nearest neighbor information by plotting the heatmap of pairwise mean squared logarithmic distance defined as
\begin{equation}
    d_\text{MSL} := \underset{\theta_{\rv{h}}, \theta_{\rv{d}}, \varphi_{\rv{d}}}{\mathbb{E}} \Big[ \left( \log \left(1 + f_r \cos \theta_i \right) - \log \left(1 + f_r' \cos \theta_i \right) \right)^2 \Big]. 
\end{equation}
From the Figure we can see that the nearest neighbor is captured effectively with the underlying distance, validating the design choice of our BRDF distributional metrics.

% \section{PCA ideal conditional generation}
% We also randomly interpolate three PCA-based MERL materials with random
% color channel permutation, it serves as another conditional generation
% approach (\cref{tbl: gen-eval-metrics-pca-ideal}). The performance
% between generated 120 materials and 120 test set is measured, where FID
% = 1.2531. We argue that due to the random selection process, which is
% intrinsically different from the proposed methods, PCA scores high marks
% on coverage-related metrics.

% % upper bound for diversity

% \begin{table*}
%   \centering
%   \rowcolors{2}{}{LightCyan}
%   \begin{tabular}{@{}lllll@{}}
%   \toprule
%    MMD & -ref & -sam & COV(↑) & 1-NNA(50\%) \\
%   \midrule
%   L1-log-cos (↓) & 0.4157 & 0.4281 & 43.33\% & 78.33\% \\
%   L1-log (↓) & 0.4475 & 0.4745 & 44.17\% & 78.33\% \\
%   L1 (↓) & 3071.2 & 1954.1 & 44.99\% & 62.08\% \\
%   \hline
%   RMSE (↓) & 0.1027 & 0.1001 & 45.83\% & 70.83\% \\
%   PSNR (↑) & 24.513 & 25.605 & 46.67\% & 70.42\% \\
%   SSIM (↑) & 0.9387 & 0.9175 & 51.67\% & 70.83\% \\
%   \bottomrule
%   \end{tabular}
%   \caption{ PCA based conditional models performance on test set.}
%   \label{tbl: gen-eval-metrics-pca-ideal}
% \end{table*}

\section{Further Synthesis Results}
\label{sec:Further Results}

\begin{figure*}
  \centering
  \includegraphics[width=\linewidth]{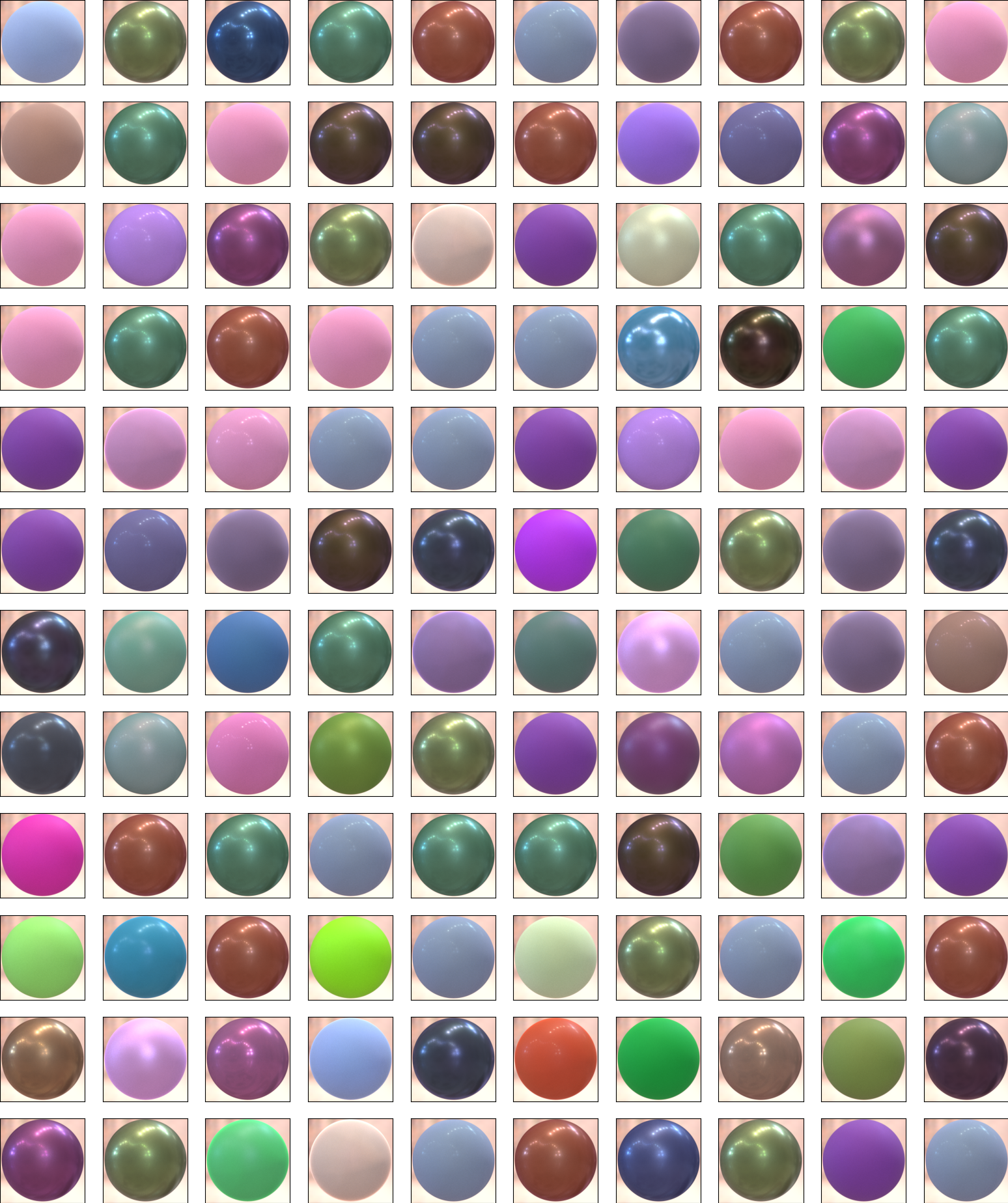}
  \caption{Synthesized materials by \ours (our method).}
  \label{fig:ours-uncond-all}
\end{figure*}

\begin{figure*}
  \centering
  \includegraphics[width=\linewidth]{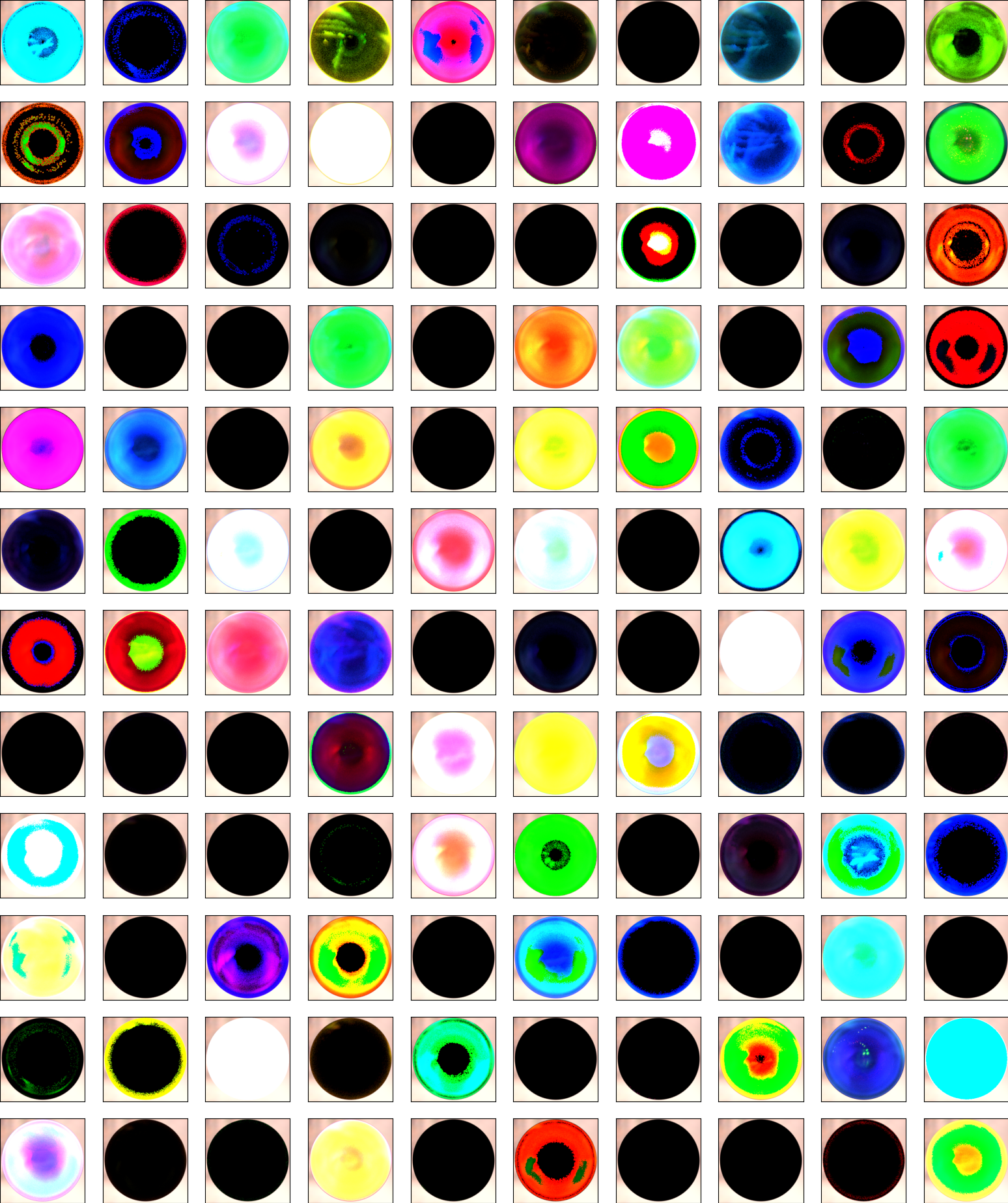}
  \caption{Synthesized materials by PCA-A.}
  \label{fig:pca-a-uncond-all}
\end{figure*}

\begin{figure*}
  \centering
  \includegraphics[width=\linewidth]{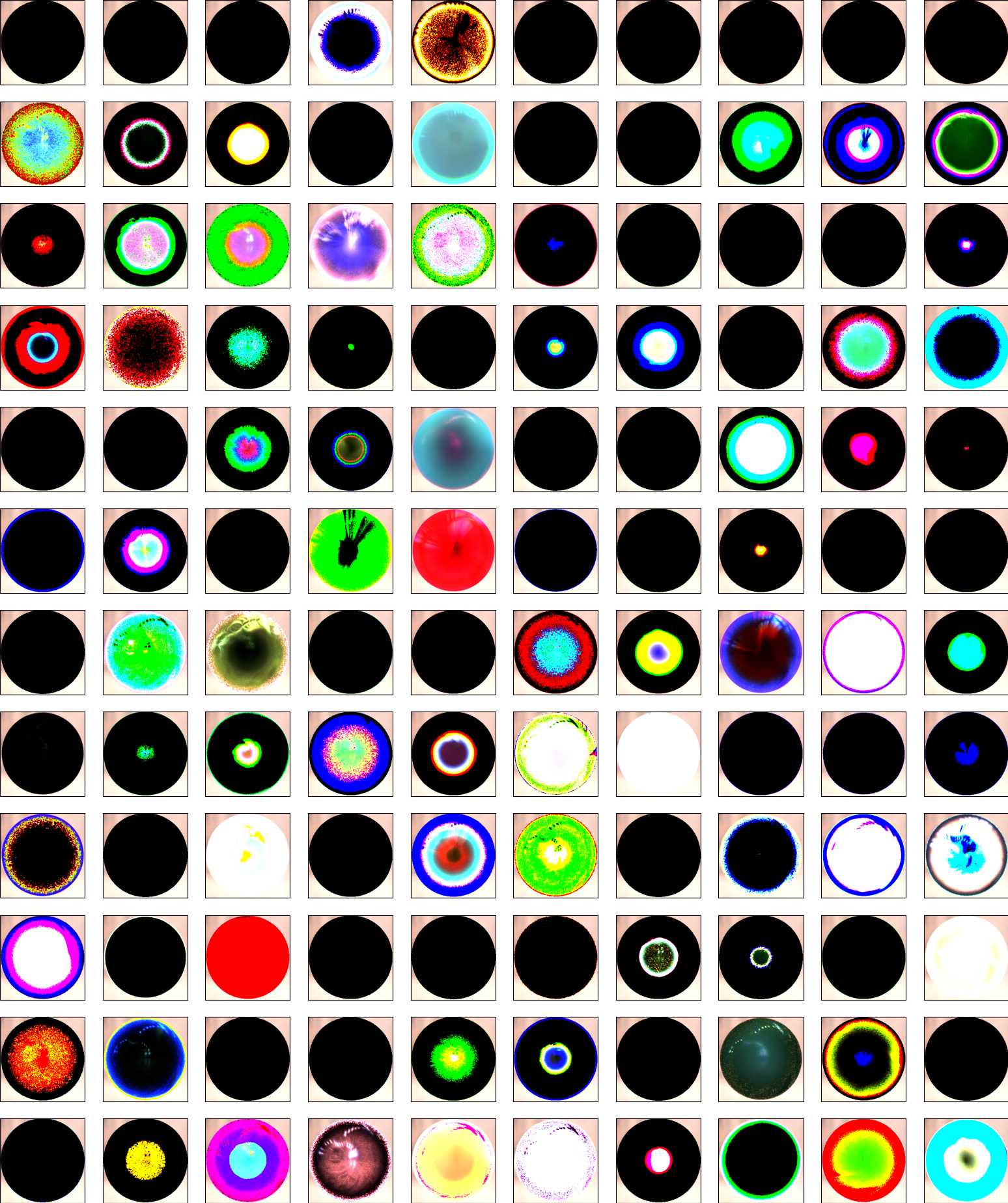}
  \caption{Synthesized materials by PCA-N.}
  \label{fig:pca-n-uncond-all}
\end{figure*}

\begin{figure*}
  \centering
  \includegraphics[width=\linewidth]{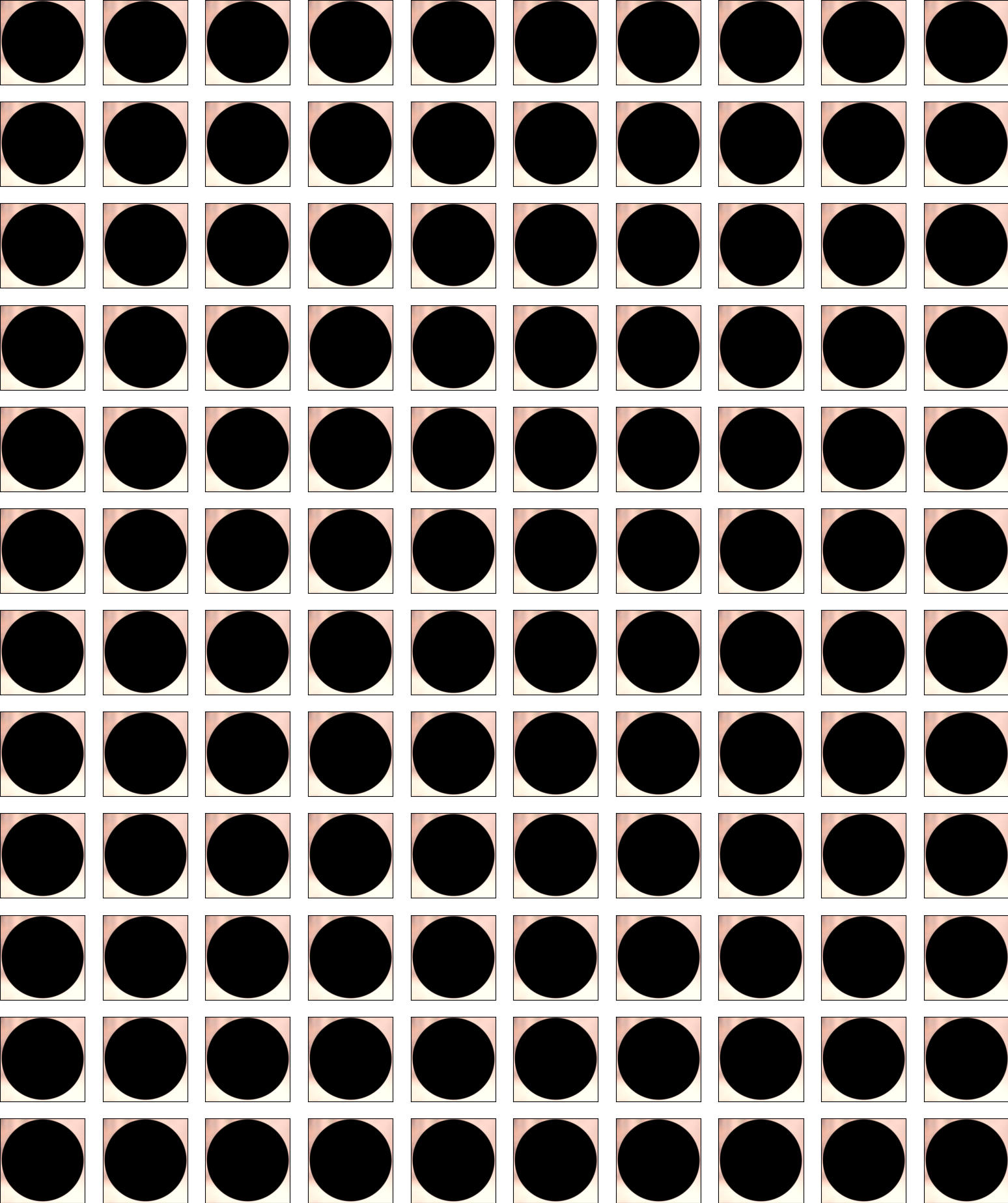}
  \caption{Synthesized materials by VAE-A.}
  \label{fig:vae-a-uncond-all}
\end{figure*}

\begin{figure*}
  \centering
  \includegraphics[width=\linewidth]{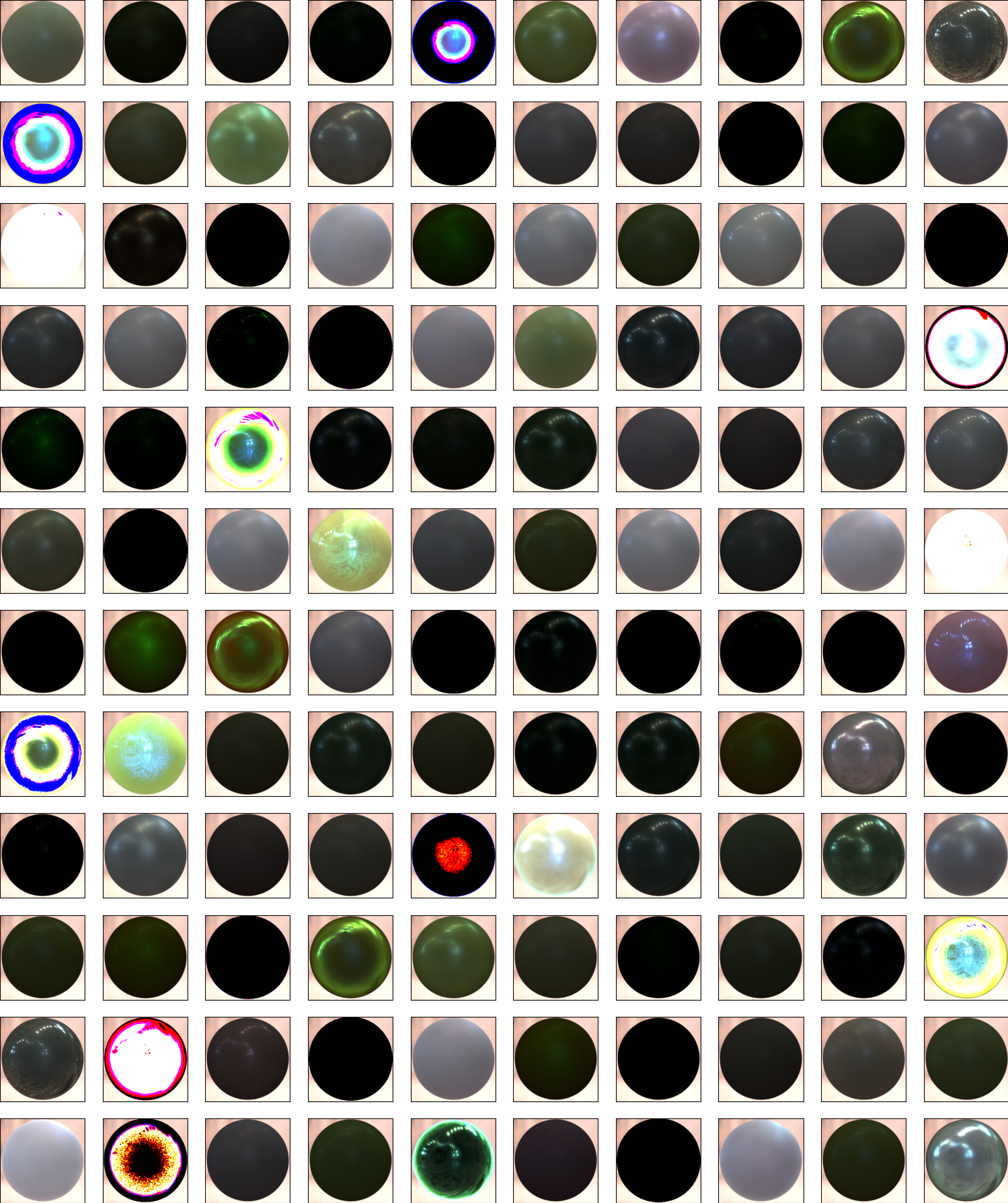}
  \caption{Synthesized materials by VAE-N.}
  \label{fig:vae-n-uncond-all}
\end{figure*}

\begin{figure*}
    \centering
    \includegraphics[width=\linewidth]{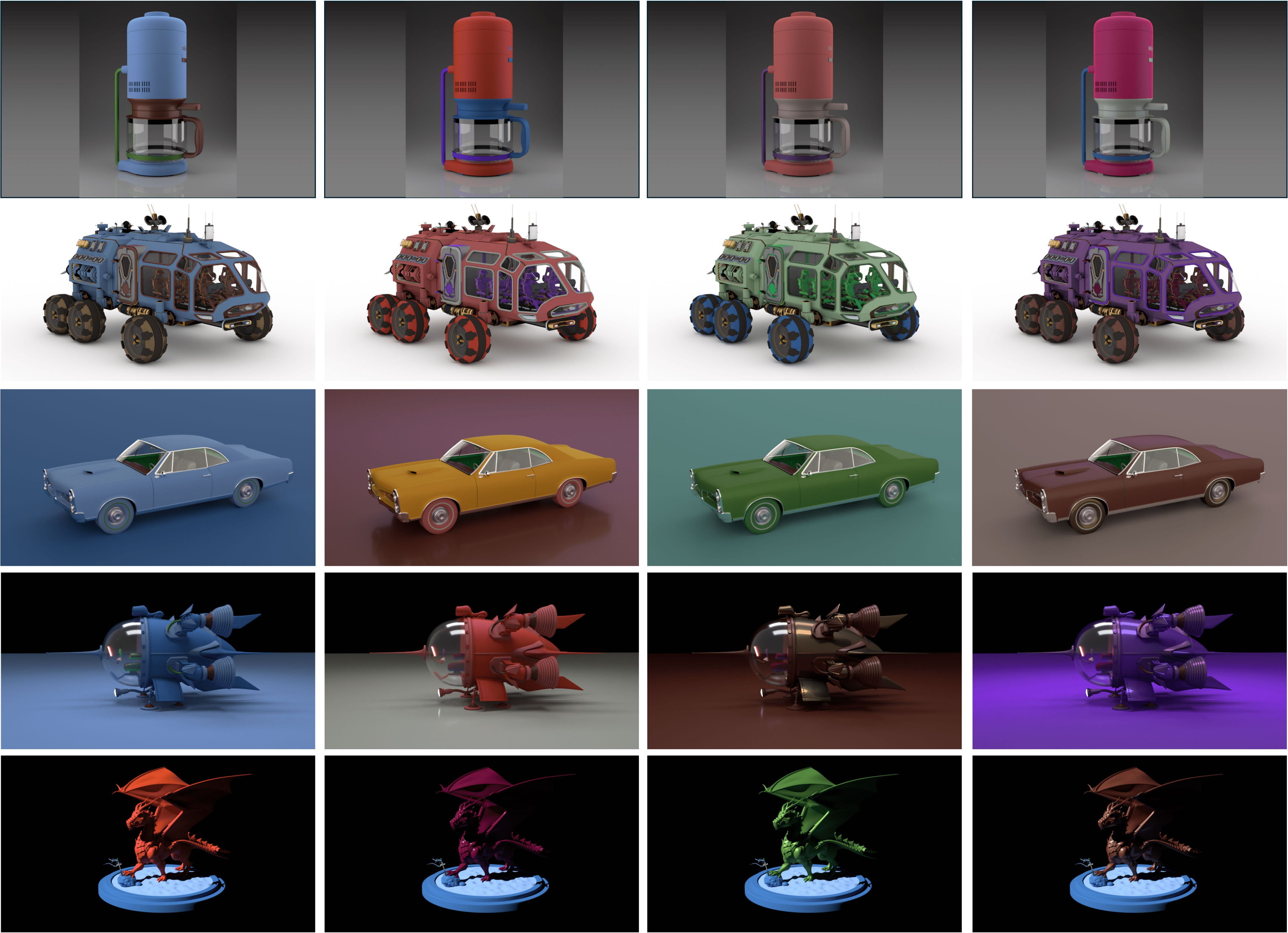}
    \caption{Renderings of different 3D models using our synthesized neural materials, highlighting the quality and diversity.}
    \label{fig:cool-models}
\end{figure*}

\begin{figure*}
    \centering
    \includegraphics[width=\linewidth]{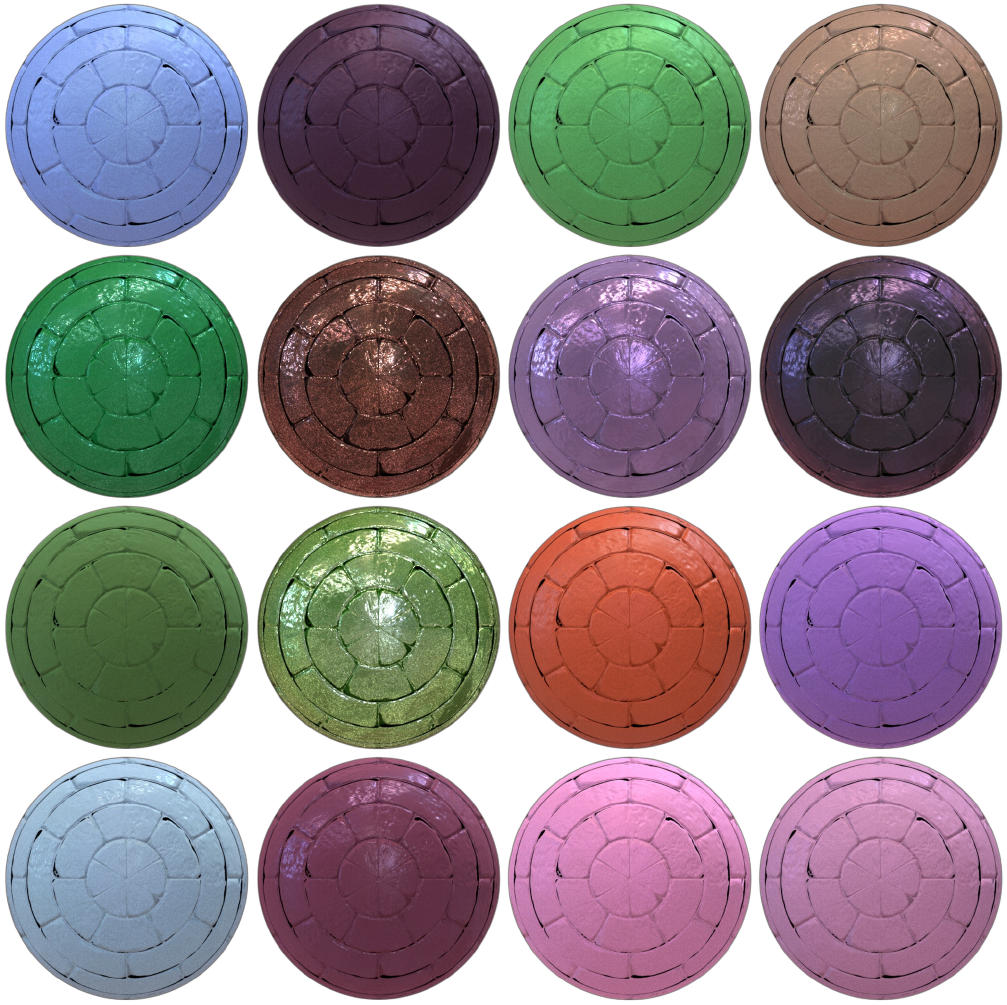}
    \caption{Our synthesized neural materials rendered with normal maps.}
    \label{fig:normal-map}
\end{figure*}

\begin{figure*}
    \centering
    \includegraphics[width=\linewidth]{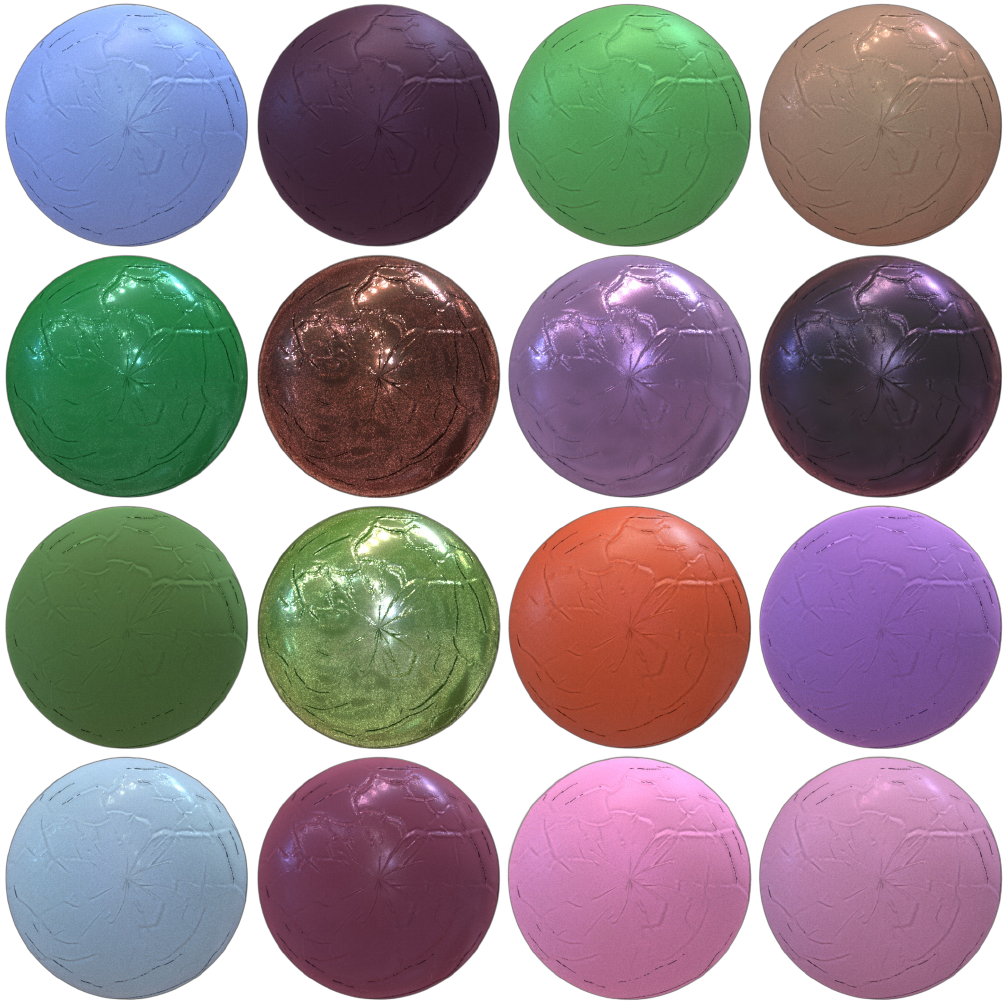}
    \caption{Our synthesized neural materials rendered with bump maps.}
    \label{fig:bump-map}
\end{figure*}

\begin{figure*}
    \centering
    \includegraphics[width=\linewidth]{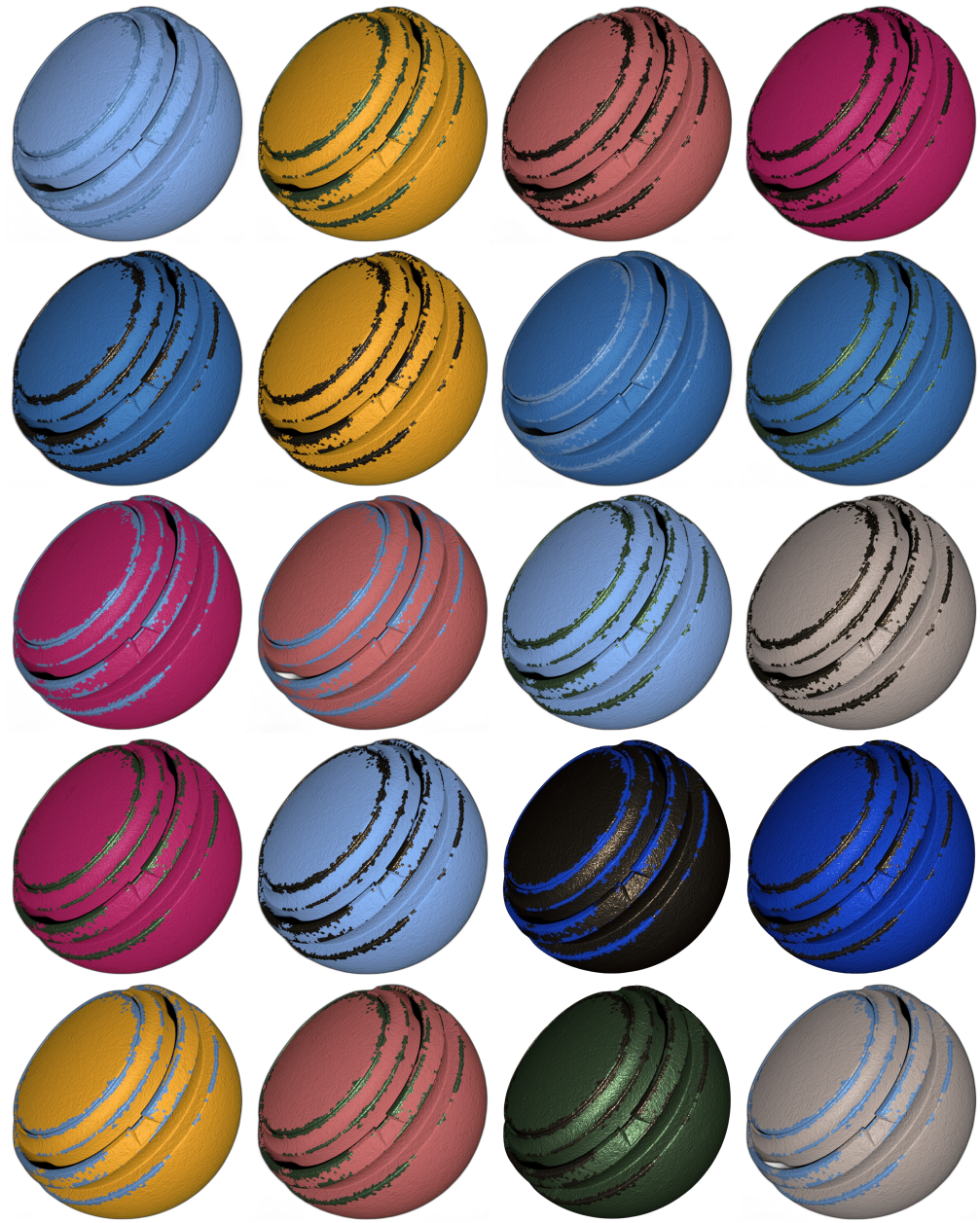}
    \caption{Our synthesized neural materials rendered in a spatially varying.}
    \label{fig:moon-rock}
\end{figure*}

\begin{figure*}
    \centering
    \includegraphics[width=\linewidth]{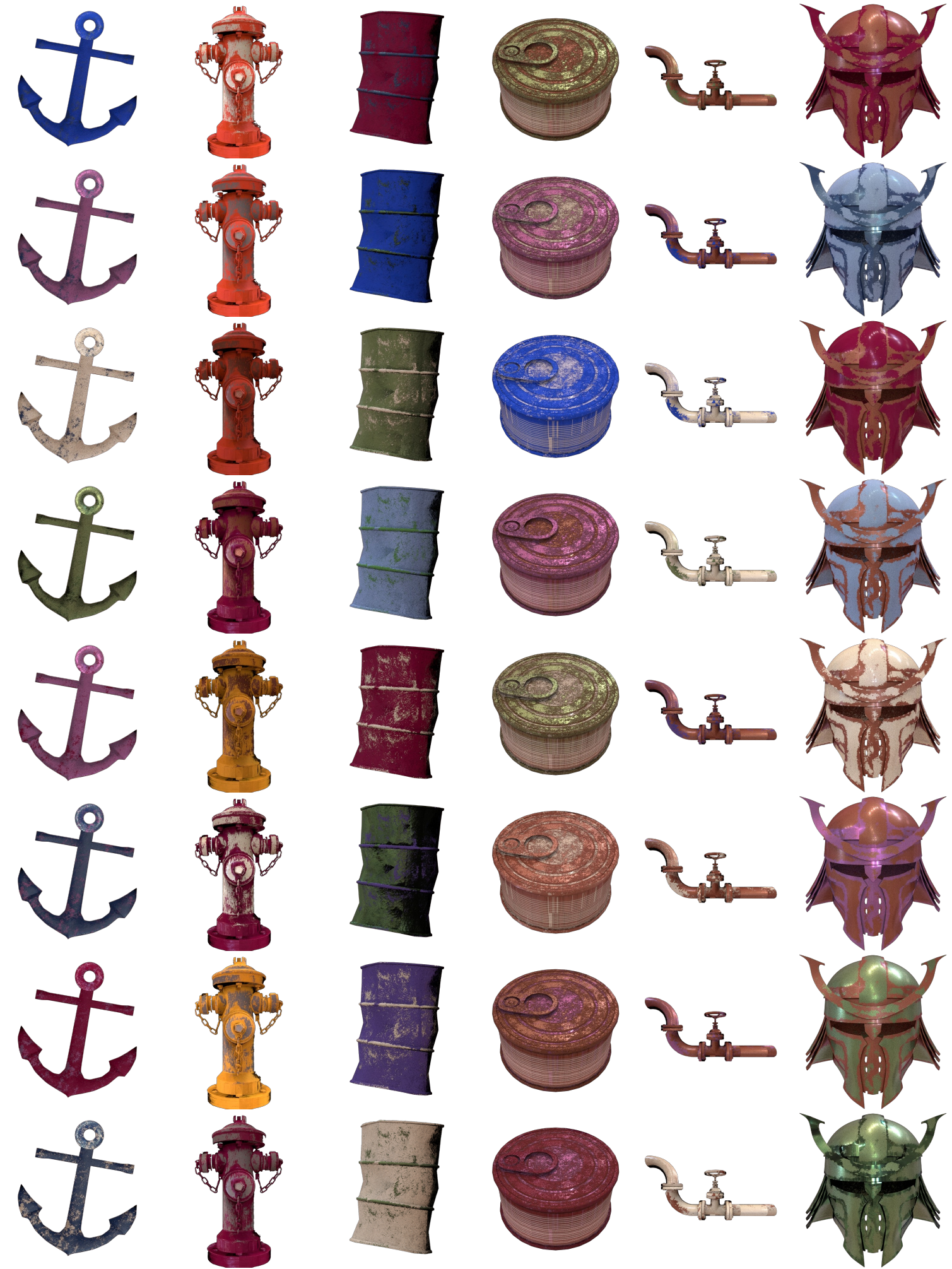}
    \caption{Renderings of different 3D models using our synthesized neural materials, highlighting the quality and diversity.}
    \label{fig:sv-materials}
\end{figure*}

For the unconditional synthesis experiment in \cref{sec:unconditional-synthesis}, further results are presented in \cref{fig:ours-uncond-all,fig:pca-a-uncond-all,fig:pca-n-uncond-all,fig:vae-a-uncond-all,fig:vae-n-uncond-all} for \ours (our method), PCA-A, PCA-N, VAE-A, and VAE-N baselines, respectively. In addition, \cref{fig:cool-models} presents the complex geometries rendered with our synthesized materials. To support richer visual effects, our synthesized materials can also be rendered with normal maps (\cref{fig:normal-map}), bump maps (\cref{fig:bump-map}) and spatially varying configurations~\cite{Jakob2022DrJit} (\cref{fig:moon-rock,fig:sv-materials}).


\section{Additional Experiments}
\label{sec:Additional Experiments}

\subsection{Statistical Analysis on MERL Dataset}
\label{sec:MERL Statistical Analysis Details}
\begin{figure}[htb]
  \centering
          \includegraphics[width=\linewidth]{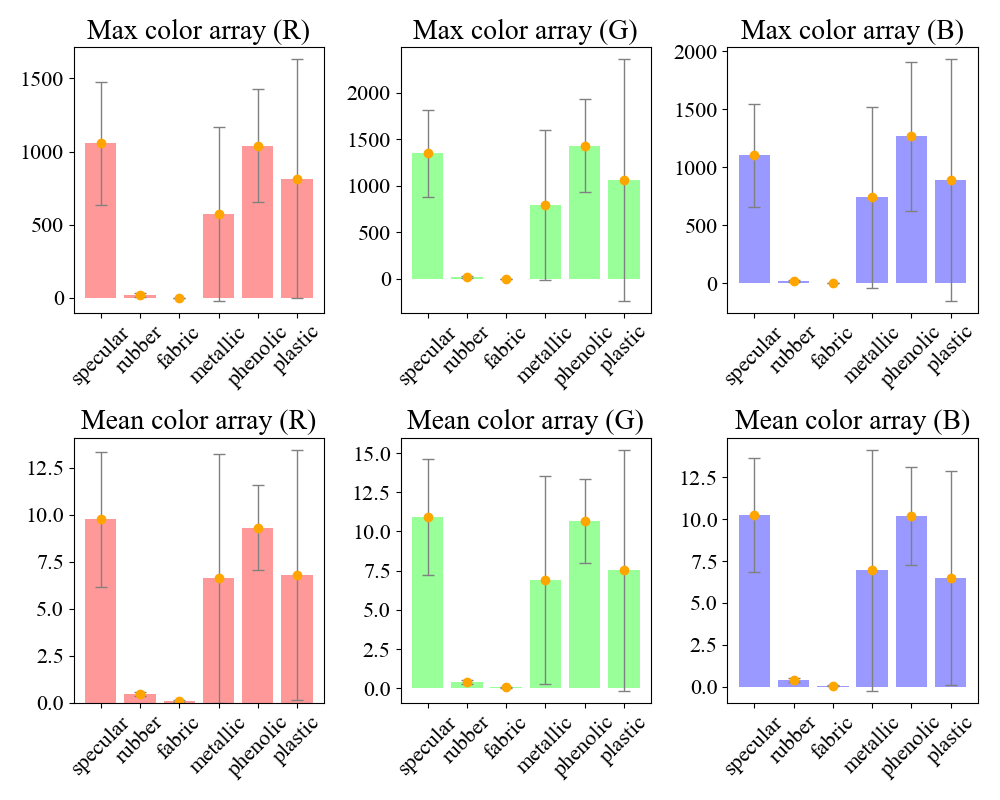}
      \caption{Per-type averaged mean and maximum reflectance.}
      \label{fig:BRDF_stat_all}
\end{figure}

\begin{figure*}[htb]
  \centering
  \begin{subfigure}{0.48\linewidth}
          \includegraphics[width=\linewidth]{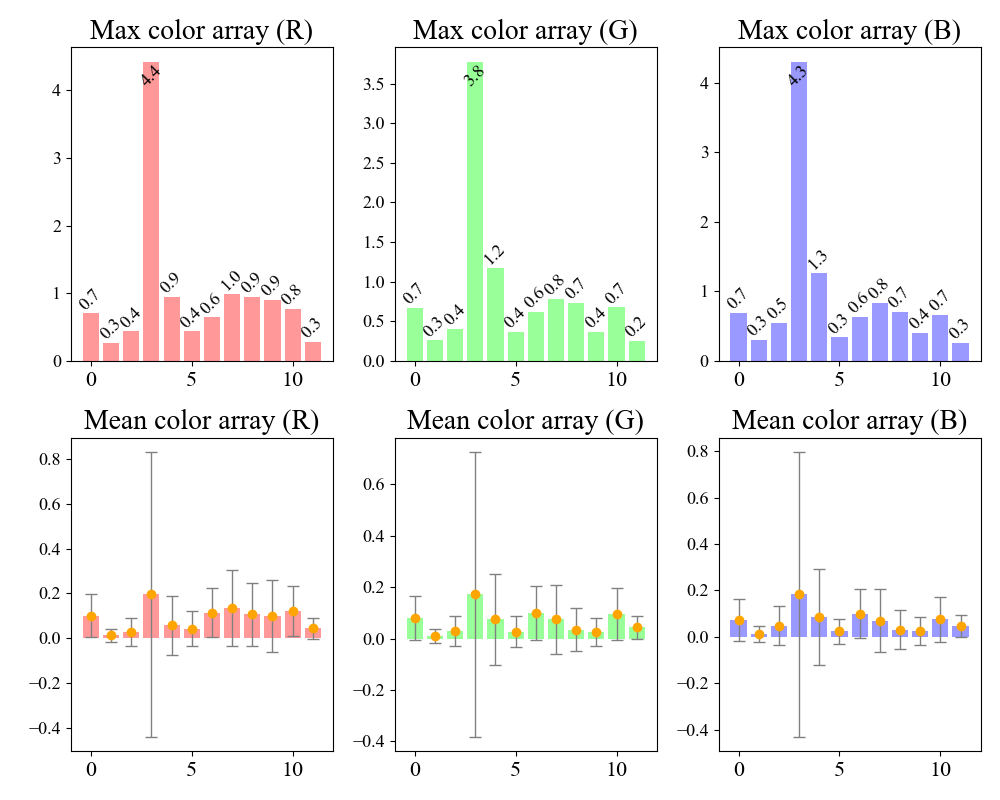}
      \caption{Fabric BRDF statistics}
  \end{subfigure}
 \begin{subfigure}{0.48\linewidth}
          \includegraphics[width=\linewidth]{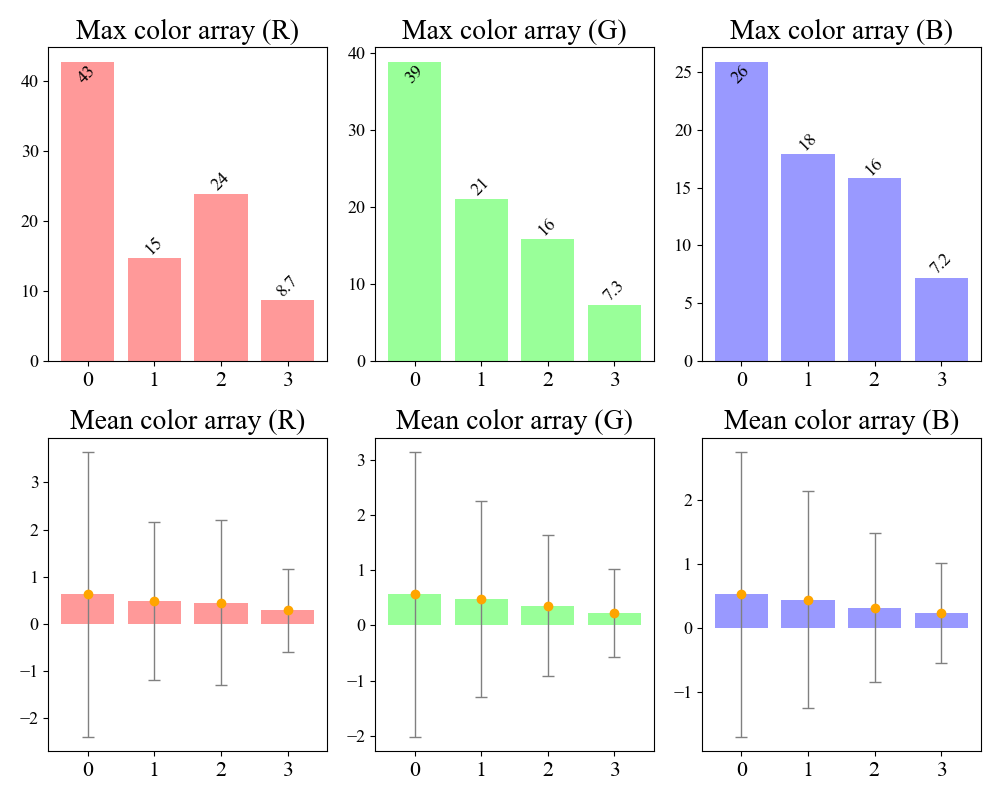}
      \caption{Rubber BRDF statistics}
  \end{subfigure}
 \begin{subfigure}{0.48\linewidth}
          \includegraphics[width=\linewidth]{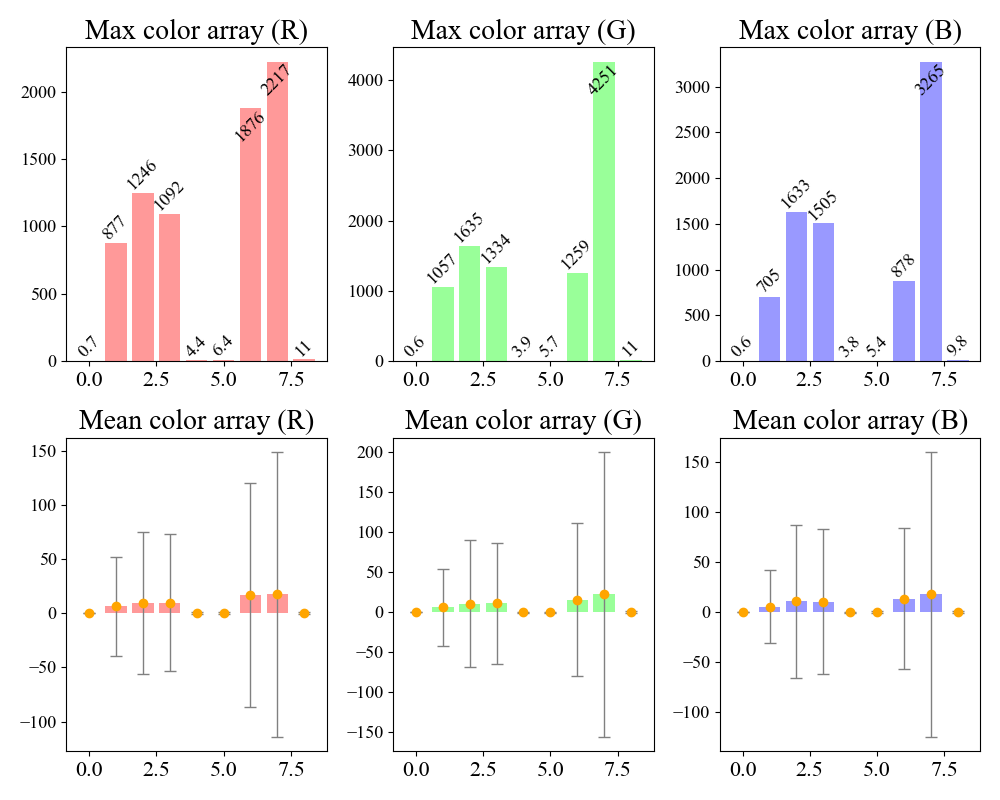}
      \caption{Plastic BRDF statistics}
  \end{subfigure}
    \begin{subfigure}{0.48\linewidth}
          \includegraphics[width=\linewidth]{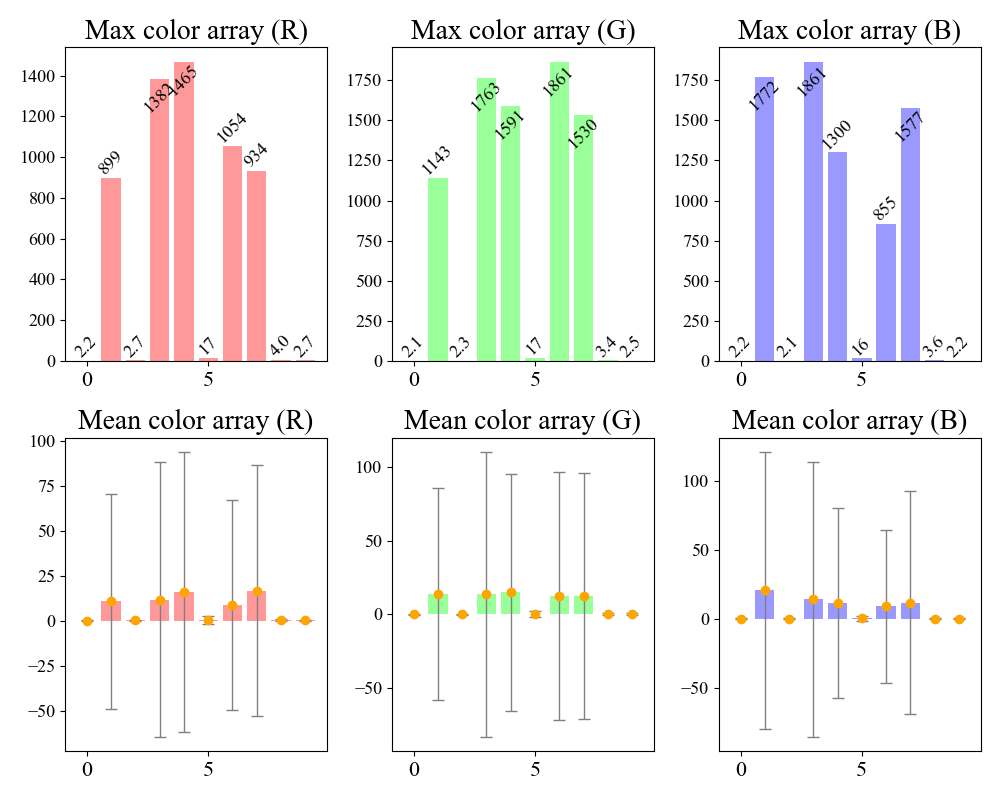}
      \caption{Metallic BRDF statistics}
  \end{subfigure}
 \begin{subfigure}{0.48\linewidth}
          \includegraphics[width=\linewidth]{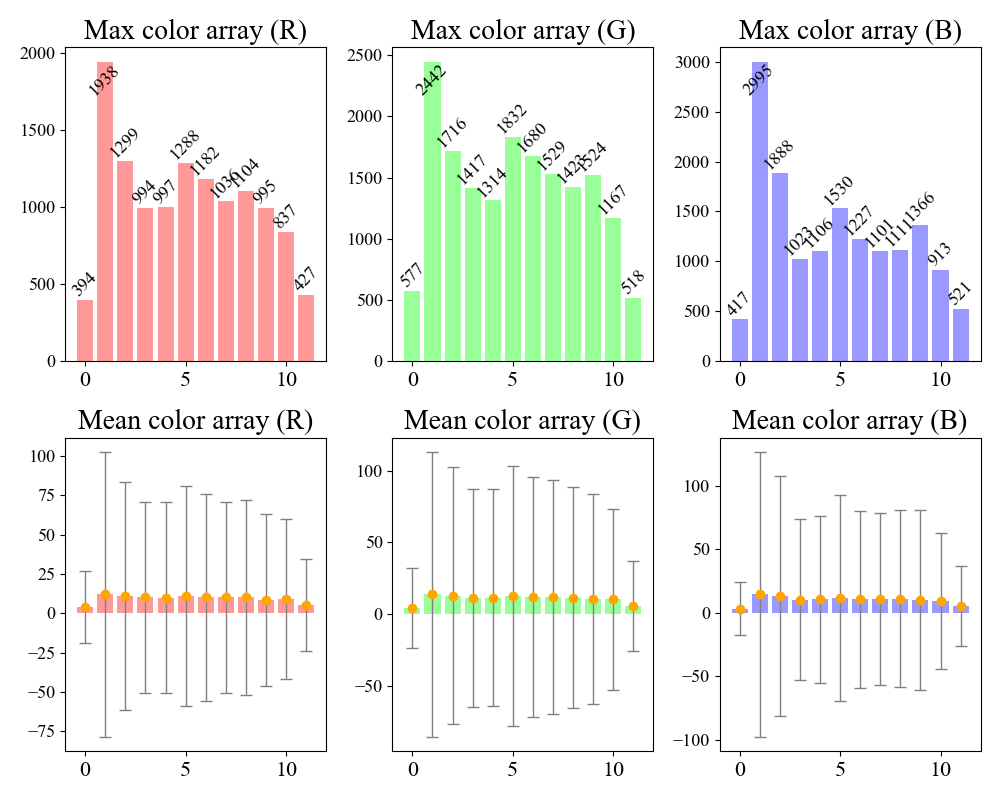}
      \caption{Phenolic BRDF statistics}
  \end{subfigure}
 \begin{subfigure}{0.48\linewidth}
          \includegraphics[width=\linewidth]{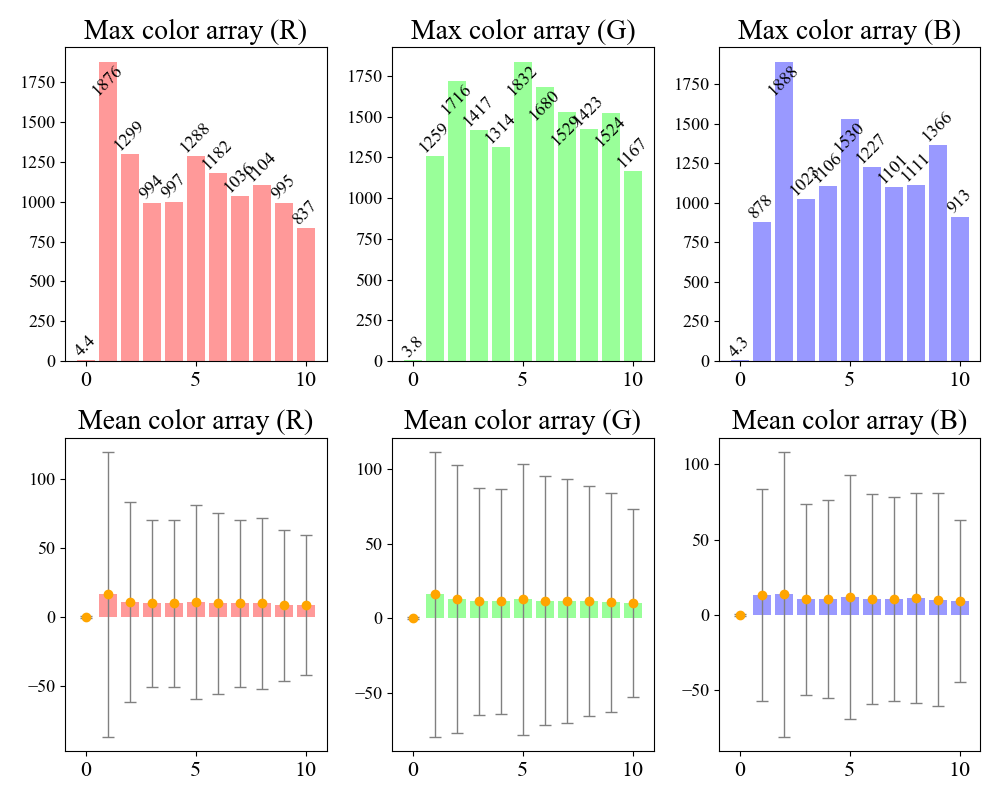}
      \caption{Specular BRDF statistics}
  \end{subfigure}
  \caption{Full MERL BRDF~\cite{Matusik2003datadriven} statistical analysis grouped by material type.}
  \label{fig:BRDF_stat_all_individual_details}
\end{figure*}

\begin{figure}[htb]
  \centering
      \includegraphics[width=\linewidth]{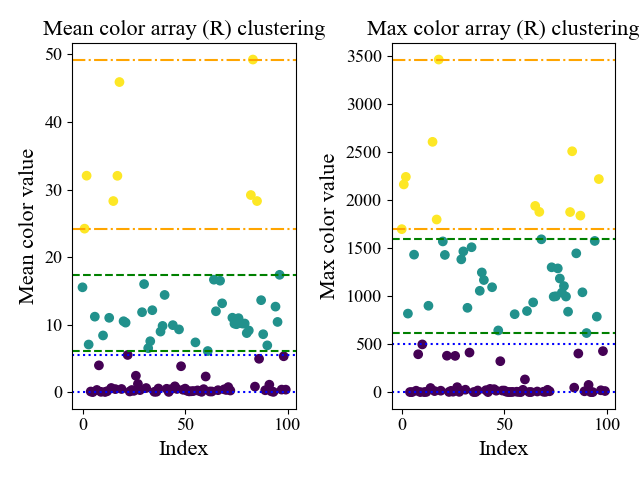}
      \caption{K-means clustering on mean (left) and maximum (right) reflectance of red channel.}
      \label{fig:k_mean_all_3_R}
\end{figure}

We perform the statistical analysis on the MERL dataset~\cite{Matusik2003datadriven}. We collect mean and maximum reflectance values in different color channels across all materials. \Cref{fig:BRDF_stat_all} demonstrates the statistical summary averaged across each material type while \cref{fig:BRDF_stat_all_individual_details} details the statistics in each individual material grouped by type. In addition, to identify the classification boundaries between different material types, in \cref{fig:k_mean_all_3_R}, we apply K-means clustering (for background see \cref{sec:k-means-clustering}) on the maximum reflectance of red channel.

\subsection{Superresolution from Low-Density BRDF}
\label{sec:superresolution-for-low-density-brdf}
We present the experiment on the BRDF reconstruction for low-density input using NBRDF~\cite{sztrajman2021nbrdf}, showing its capability of superresolution. In the experiment, when fitting the NBRDF to a MERL material~\cite{Matusik2003datadriven}, we downsample the input \(\theta_h, \theta_d, \phi_d\) by a factor of $x=1,2,4,8,16$ in all three angles. Since the original sampling density is \(90 \times 90 \times 180\), after downsampling, the number of input samples is
\[
(1+\lfloor \frac{89}{x} \rfloor) \times (1+\lfloor \frac{89}{x} \rfloor) \times (1+ \lfloor \frac{179}{x} \rfloor).
\]

For comparison we develop a baseline model, which just evaluates the BRDF according to the nearest neighbor. On the other
hand, NBRDF model is trained over this downsampled data. Under the same scene
setting, we compare SSIM of rendered images using the
full-density groundtruth, the baseline, and the NBRDF model in \cref{tbl: super-res}.
\begin{table}
  \centering
   \rowcolors{2}{}{gray!20}
  \begin{tabular}{@{}lcll@{}}
  \toprule
  \# samples & \(x\) & baseline & NBRDF \\
  \midrule
  \(90^2 \times 180\) & 1x & \textbf{1 ± 0} & 0.9987 ± 7e-6 \\
  \(23^2 \times 45\) & 4x & \textbf{0.996 ± 9.3e-5} & \textbf{0.996} ± 0.00012 \\
  \(12^2 \times 23\) & 8x & 0.984 ± 0.001 & \textbf{0.9934 ± 1.475e-4} \\
  \(6^2 \times 12\) & 16x & 0.915 ± 0.024 & \textbf{0.962 ± 0.0067} \\
  \(4^2 \times 8\) & 24x & 0.823 ± 0.0527 & \textbf{0.896 ± 0.02614} \\
  \(3^2 \times 6\) & 32x & 0.745 ± 0.07467 & \textbf{0.831 ± 0.04942} \\
  \bottomrule
  \end{tabular}
  \caption{Low-density reconstruction comparison (SSIM)}
  \label{tbl: super-res}
\end{table}

From the Table, we see that the NBRDF model exhibits significantly better reconstruction capability
over the MERL dataset.
This capability might be useful in the scenarios where high-resolution BRDF collection
is not available.

% \clearpage
% \input{aaai26/ReproducibilityChecklist}

\end{document}